\documentclass[prd,aps,showpacs,epsf,floats]{revtex4}
\usepackage{amssymb}
\usepackage{amsfonts}
\usepackage{amsmath}

\usepackage{graphicx}

\usepackage{color}

\begin{document}

\title{Quantifying the Performance of Quantum Codes}
\author{Carlo Cafaro$^{1}$, Sonia L'Innocente$^{2}$, Cosmo Lupo$^{3}$ and
Stefano Mancini$^{4,5}$}
\affiliation{$^{1,3,4}$School of Science and Technology, Physics Division, University of
Camerino, I-62032 Camerino, Italy\\
$^{2}$School of Science and Technology, Mathematics Division, University of
Camerino, I-62032 Camerino, Italy\\
$^{5}$INFN, Sezione di Perugia, I-06123 Perugia, Italy}

\begin{abstract}
We study the properties of error correcting codes for noise models in
the presence of asymmetries and/or correlations by means of the entanglement
fidelity and the code entropy. First, we consider a dephasing
Markovian memory channel and characterize the performance of both a
repetition code and an error avoiding code ($\mathcal{C}_{RC}$ and 
$\mathcal{C}_{DFS}$, respectively) in terms of the entanglement fidelity. We
also consider the concatenation of such codes ($\mathcal{C}_{DFS}\circ 
\mathcal{C}_{RC}$) and show that it is especially advantageous in the regime
of partial correlations. Finally, we characterize the effectiveness of the
codes $\mathcal{C}_{DFS}$, $\mathcal{C}_{RC}$ and 
$\mathcal{C}_{DFS}\circ \mathcal{C}_{RC}$ by means of the code entropy and
find, in particular, that the effort required for recovering such codes
decreases when the error probability decreases and the memory parameter
increases. Second, we consider both symmetric and asymmetric depolarizing
noisy quantum memory channels and perform quantum error correction via the
five qubit stabilizer code $\mathcal{C}_{\left[ \left[ 5,1,3\right] %
\right] }$. We characterize this code by means of the entanglement fidelity
and the code entropy as function of the asymmetric error probabilities and
the degree of memory. Specifically, we uncover that while the asymmetry in 
the depolarizing errors does not affect the entanglement fidelity of the
five qubit code, it becomes a relevant feature when the code entropy is 
used as a performance quantifier.
\end{abstract}

\maketitle

\section{Introduction}

Quantum error correction (QEC) is a theoretical scheme developed in
quantum computing to defend quantum coherence against
environmental noise. There are different methods for preserving quantum
coherence. One possible technique exploits the redundancy in encoding
information. This scheme is known as "quantum error correcting codes"
(QECCs). For a comprehensive introduction to QECCs, we refer to \cite{gotty1}%
. Within such scheme, information is encoded in linear subspaces (codes) of
the total Hilbert space in such a way that errors induced by the interaction
with the environment can be detected and corrected. The QECC approach may be
interpreted as an active stabilization of a quantum state in which, by
monitoring the system and conditionally carrying on suitable operations, one
prevents the loss of information. In detail, the errors occur on a qubit
when its evolution differs from the ideal one. This happens by
interaction of the qubit with an environment. Another possible approach is
the so-called noiseless codes (also known as decoherence free subspaces
(DFSs) or error avoiding codes). For a comprehensive introduction to DFSs,
we refer to \cite{lidar1}. It turns out that for specific open quantum
systems (noise models in which all qubits can be considered symmetrically
coupled with the same environment), it is possible to design states that are
hardly corrupted rather than states that can be easily corrected (in this
sense, DFSs are complementary to QECCs). In other words, it is possible to
encode information in linear subspaces such that the dynamics restricted to
such subspaces is unitary. This implies that no information is lost and
quantum coherence is maintained. DFS is an example of passive stabilization
of quantum information.

The formal mathematical description of the qubit-environment interaction is
usually given in terms of quantum channels. Quantum error correction is
usually developed under the assumption of i.i.d. (identically and
independently distributed) errors. These error models are characterized by
memoryless channels $\Lambda $ such that $n$-channel uses is
given by $\Lambda ^{\left( n\right) }=\Lambda ^{\otimes n}$. In such cases
of complete independent decoherence, qubits interact with their own
environments which do not interact with each other. However, in many
physical situations, qubits do interact with a common environment which
unavoidably introduces correlations in the noise. For instance, there are
situations where qubits in a ion trap set-up are collectively coupled to
their vibrational modes \cite{garg96}. In other situations, different qubits
in a quantum dot design are coupled to the same lattice, thus interacting
with a common thermal bath of phonons \cite{loss98}. The exchange of bosons
between qubits causes spatial and temporal correlations that violate the
condition of error independence \cite{hwang01}. Memory effects then
arise among channel uses with the consequence that $\Lambda ^{\left(
n\right) }\neq \Lambda ^{\otimes n}$. Recent studies try to characterize the
effect of correlations on the performance of QECCs \cite{clemens04,
klesse05, shabani08, d'arrigo08, carlo-PLA, carloPRA}. It appears that
correlations may have negative \cite{klesse05} or positive \cite{shabani08}
impact on QECCs depending on the features of the error model being
considered.

A part from being correlated, noise errors may also be asymmetric. Most of
the quantum computing devices \cite{astafiev04} are characterized by
relaxation times ($\tau _{\text{relaxation}}$) that are one-two orders of
magnitude larger than the corresponding dephasing times ($\tau _{\text{%
dephasing}}$). Relaxation leads to both bit-flip and phase-flip errors,
whereas dephasing (loss of phase coherence) only leads to
phase-flip errors. Such asymmetry between $\tau _{\text{relaxation}}$ and $%
\tau _{\text{dephasing}}$ translates to an asymmetry in the occurrence
probability of bit-flip ($p_{X}$) and phase-flip errors ($p_{Z}$). The ratio 
$\frac{p_{Z}}{p_{X}}$ is known as the channel asymmetry. Quantum error
correction schemes should be designed in such a way that no resources (time
and qubits) are wasted in attempting to detect and correct errors that may
be relatively unlikely to occur. Quantum codes should be designed in order
to exploit this asymmetry and provide better performance by neglecting the
correction of less probable errors \cite{ioffe07, evans07, stephens08}.
Indeed, examples of efficient quantum error-correcting codes taking
advantage of this asymmetry are given by families of codes of the
Calderbank-Shor-Steane (CSS) type (asymmetric stabilizer CSS\ codes)
\cite{sarvepalli08, aly08}.

Devising new good quantum codes for independent, correlated and asymmetric
noise models is a highly non trivial problem. However, it is usually
possible to manipulate in a smart way existing codes to construct
new ones suitable for more general error models and with higher performances 
\cite{gotty2, calder1}. Concatenation is perhaps one of the most successful
quantum coding tricks employed to produce new codes from old ones. The
concept of concatenated codes was first introduced in classical error
correcting schemes by Forney \cite{forney}. Roughly speaking, concatenation
is a method of combining two codes (an inner and an outer code) to form a
larger code. In classical error correction, Forney has carried on extensive
studies of concatenated codes, how to choose the inner and outer codes, and
what error probability threshold values can be achieved. The first
applications of concatenated codes in quantum error correction appear in 
\cite{gotty3, knill1}. In the quantum setting, concatenated codes play a key
role in fault tolerant quantum computation and in constructing good
degenerate quantum error correcting codes.

Obviously enough, the choice of a code depends on the error model.
However, in selecting a code rather than another, one should take into
account their performance on the error model. Since there is no single
quantity holding all information on a code, different quantum code
performance quantifiers have appeared into the QEC literature \cite%
{nielsen-book}. For instance, in \cite{nielsen-fidel} it is suggested that
the entanglement fidelity \cite{schumi} is the important quantity to
maximize in schemes for quantum error correction. In \cite{kribs}, it is
proposed that the code entropy is an important quantitative measure
of the quantum correction schemes by determining a hierarchical structure in
the set of protected spaces \cite{dark}.

Motivated by all the above mentioned considerations, in this work we study
the properties of error correcting codes for noise models in the
presence of asymmetries and/or correlations by means of the entanglement
fidelity and the code entropy. First, we consider a dephasing
Markovian memory channel and characterize the performance of both a 
repetition code and an error avoiding code ($\mathcal{C}_{RC}$ and 
$\mathcal{C}_{DFS}$, respectively) in terms of the entanglement fidelity. We
also consider the concatenation of such codes ($\mathcal{C}_{DFS}\circ 
\mathcal{C}_{RC}$) and show that it is especially advantageous in the regime
of partial correlations. Finally, we characterize the effectiveness of the
codes $\mathcal{C}_{DFS}$, $\mathcal{C}_{RC}$ and $%
\mathcal{C}_{DFS}\circ \mathcal{C}_{RC}$ by means of the code entropy. We
show that the effort required for recovering such codes decreases when the
error probability decreases and/or the memory parameter increases. Second,
we consider an asymmetric depolarizing noisy quantum memory channel and
perform quantum error correction via the five qubit stabilizer code 
$\mathcal{C}_{\left[ \left[ 5,1,3\right] \right] }$ . We characterize this
code by means of the entanglement fidelity and the code entropy as function
of the asymmetric error probabilities and the degree of memory.
Specifically, we uncover that while the asymmetry in the depolarizing 
errors does not affect the entanglement fidelity of the five qubit code, it
becomes a relevant feature when quantifying the code performance by means of
the code entropy.

The layout of this article is as follows. In Section II, we present few
theoretical tools employed in the rest of the article. In particular, we
emphasize the conceptual and operational meanings of the entanglement
fidelity and code entropy, the two code performance quantifiers employed
here. In Section III, we study a dephasing Markovian memory channel
and characterize the performance of both a repetition code and
an error avoiding code ($\mathcal{C}_{RC}$ and $\mathcal{C}_{DFS}$,
respectively) in terms of the entanglement fidelity. We also consider the
concatenation of such codes ($\mathcal{C}_{DFS}\circ \mathcal{C}_{RC}$) and
show that it is especially advantageous in the regime of partial
correlations. Finally, we characterize the effectiveness of the
codes $\mathcal{C}_{DFS}$, $\mathcal{C}_{RC}$ and $\mathcal{C}%
_{DFS}\circ \mathcal{C}_{RC}$ by means of the code entropy. In Section IV,
we analyze both symmetric and asymmetric depolarizing noisy quantum memory
channels and perform quantum error correction via the five qubit
stabilizer code $\mathcal{C}_{\left[ \left[ 5,1,3\right] \right] }$. We
characterize this code by means of the entanglement fidelity and the code
entropy as function of the asymmetric error probabilities and the degree of
memory. We find, among other things, that while asymmetric depolarizing
errors do not affect the entanglement fidelity of the five qubit code, they
do affect its code entropy. A summary of the uncovered results and our
comments on the usefulness of both the entanglement fidelity and code
entropy as suitable code performance quantifiers for the error models
considered appear in Section V.

\section{Preliminaries}

In this Section, we briefly describe some basic features of quantum error
correcting codes and decoherence free subspaces. Finally, we point out the
conceptual and operational meanings of the entanglement fidelity and code
entropy, the two code performance quantifiers employed in this work.

\subsection{On quantum error correction schemes}

\subsubsection{Quantum error correcting codes}

The formal mathematical description of a system-environment interaction is
usually given in terms of quantum channels \cite{nielsen-book}. When a
quantum system in a state $\rho $ (belonging to the set of density
operators $\mathcal{\sigma }(\mathcal{H})$ defined over the Hilbert space $%
\mathcal{H}$ of dimension $d$ associated to the system) is exposed to the
interaction with the environment, the result is the noisy state describable
by the action of a completely positive and trace preserving map 
$\Lambda :\mathcal{\sigma }(\mathcal{H})\rightarrow \mathcal{\sigma }(%
\mathcal{H})$ with,%
\begin{equation}
\Lambda \left( \rho \right) =\sum_{k}A_{k}\rho A_{k}^{\dagger }\text{.}
\label{channel}
\end{equation}%
The operators $A_{k}:\mathcal{H}\rightarrow \mathcal{H}$ (Kraus operators)
define the errors affecting the system's state $\rho $. It is well known 
\cite{nielsen-book} that a standard quantum error correction scheme is
characterized by a subspace $\mathcal{C}\subseteq $ $\mathcal{H}$ such that
for some Hermitian and positive matrix of complex scalars $\Gamma =\left(
\gamma _{lm}\right) $ the corresponding projection operator $P_{\mathcal{C}}$
fulfills the error correction condition,%
\begin{equation}
P_{\mathcal{C}}A_{l}^{\dagger }A_{m}P_{\mathcal{C}}=\gamma _{lm}P_{\mathcal{C%
}}\text{,}  \label{ecc}
\end{equation}%
for $l$, $m=1$,..., $d^{2}$. The matrix $\Gamma $ is also known as the error
correction matrix. The action of the correctable error operators $\{A\}_{%
\text{correctable}}\subseteq \{A_{k}\}$ on vectors spanning $\mathcal{C}$
(codewords) allows to define the (completely positive and trace preserving)
recovery map $\mathcal{R}\leftrightarrow \left\{ R_{l}\right\} $. The
composition of the recovery map $\mathcal{R}$ with the error map $\mathcal{A}
$ gives the recovered channel,%
\begin{equation}
\Lambda _{\text{rec.}}\left( \rho \right) \overset{\text{def}}{=}\left( 
\mathcal{R}\circ \Lambda \right) \left( \rho \right) =\sum_{l}\sum_{k}\left(
R_{l}A_{k}\right) \rho \left( R_{l}A_{k}\right) ^{\dagger }\text{.}
\end{equation}%
For a comprehensive introduction to QECCs, we refer to \cite{gotty1}.

\subsubsection{Decoherence Free Subspaces}

Following \cite{lidar1}, we mention few relevant properties of DFSs.
Consider the dynamics of a closed system composed of a quantum system $%
\mathcal{Q}$ coupled to a bath $\mathcal{B}$. The unitary evolution of the
closed system is described by the combined system-bath Hamiltonian $H_{\text{%
tot}}$, $H_{\text{tot}}=H_{\mathcal{Q}}\otimes I_{\mathcal{B}}+H_{\mathcal{B}%
}\otimes I_{\mathcal{Q}}+H_{\text{int}}$, $H_{\text{int}}=\sum_{\alpha
}E_{\alpha }\otimes B_{\alpha }$. The operator $H_{\mathcal{Q}}$ ($H_{%
\mathcal{B}}$) is the system (bath) Hamiltonian, $I_{\mathcal{Q}}$ ($I_{%
\mathcal{B}}$) is the identity operator of the system (bath), $E_{\alpha }$
are the error generators acting solely on $\mathcal{Q}$ while $B_{\alpha }$
act on the bath. The last term in $H_{\text{tot}}$ is the interaction
Hamiltonian.

A subspace $\mathcal{H}_{\text{DFS}}$ of the total system Hilbert space $%
\mathcal{H}$ is a decoherence free subspace if and only if: i) $%
E_{\alpha }\left\vert \psi \right\rangle =c_{\alpha }\left\vert \psi
\right\rangle $ with $c_{\alpha }\in 
\mathbb{C}
$, for all states $\left\vert \psi \right\rangle $ spanning $\mathcal{H}_{%
\text{DFS}}$, and for every error operator $E_{\alpha }$ in $H_{\text{int}}$%
. In other words, all basis states spanning $\mathcal{H}_{\text{DFS}}$ are
degenerate eigenstates of all the error generators $E_{\alpha }$; ii) $%
\mathcal{Q}$ and $\mathcal{B}$ are initially decoupled; iii) $H_{\mathcal{Q}%
}\left\vert \psi \right\rangle $ has no overlap with states in the subspace
orthogonal to $\mathcal{H}_{\text{DFS}}$. To establish a direct link between
QECCs and DFSs, it is more convenient to present an alternative formulation
of DFSs in terms of the Kraus operator sum representation. Within such
description, the evolution of the system $\mathcal{Q}$ density matrix is
written as, $\rho _{\mathcal{Q}}\left( t\right) =Tr_{\mathcal{B}}\left[
U\left( \rho _{\mathcal{Q}}\otimes \rho _{\mathcal{B}}\right) U^{\dagger }%
\right] =\sum_{k}A_{k}\rho _{\mathcal{Q}}\left( 0\right) A_{k}^{\dagger }$,
where $U=e^{-iH_{\text{tot}}t}$ is the unitary evolution operator for the
system-bath closed system and the initial bath density matrix $\rho _{%
\mathcal{B}}$ equals $\sum_{n}\lambda _{n}\left\vert n\right\rangle
\left\langle n\right\vert $. The Kraus operators $A_{k}$ (satisfying the
normalization condition) are given by $A_{k}=\sqrt{\lambda _{n}}%
\left\langle m|U|n\right\rangle $ with $\sum_{k}A_{k}^{\dagger }A_{k}=I_{%
\mathcal{Q}}$ with $k=\left( n\text{, }m\right) $ and where $\left\vert
m\right\rangle $ and $\left\vert n\right\rangle $ are bath states. It turns
out that a $N_{\text{DFS}}$-dimensional subspace $\mathcal{H}_{\text{DFS}}$
of $\mathcal{H}$ is a DFS if and only if all Kraus operators have an
identical unitary representation (in the basis where the first $N_{\text{DFS}%
}$ states span $\mathcal{H}_{\text{DFS}}$) upon restriction to it, up to a
multiplicative constant,%
\begin{equation}
A_{k}=\left( 
\begin{array}{cc}
g_{k}U_{\mathcal{Q}}^{\left( \text{DFS}\right) } & 0 \\ 
0 & \bar{A}_{k}%
\end{array}%
\right) \text{,}  \label{3}
\end{equation}%
where $g_{k}=\sqrt{n}\left\langle m|U_{\text{c}}|n\right\rangle $ and $U_{%
\text{c}}=e^{-iH_{\text{c}}t}$ with $H_{\text{c}}=H_{\mathcal{B}}+H_{\text{%
int}}$. Furthermore, $\bar{A}_{k}$ is an arbitrary matrix that acts on $%
\mathcal{H}_{\text{DFS}}^{\perp }$ (with $\mathcal{H}=\mathcal{H}_{\text{DFS}%
}\oplus \mathcal{H}_{\text{DFS}}^{\perp }$) and may cause decoherence there; 
$U_{\mathcal{Q}}^{\left( \text{DFS}\right) }$ is $U_{\mathcal{Q}}$
restricted to $\mathcal{H}_{\text{DFS}}$. Now recall that in ordinary QECCs,
it is possible to correct the errors induced by a given set of Kraus
operators $\left\{ A_{k}\right\} $ if and only if,%
\begin{equation}
R_{r}A_{k}=\left( 
\begin{array}{cc}
\lambda _{rk}I_{\mathcal{C}} & 0 \\ 
0 & B_{rk}%
\end{array}%
\right) \text{,}  \label{1}
\end{equation}%
$\forall r$ and $k$, or equivalently,%
\begin{equation}
A_{k}^{\dagger }A_{k^{\prime }}=\left( 
\begin{array}{cc}
\gamma _{kk^{\prime }}I_{\mathcal{C}} & 0 \\ 
0 & \bar{A}_{k}^{\dagger }\bar{A}_{k^{\prime }}%
\end{array}%
\right) \text{,}  \label{2}
\end{equation}%
where $\left\{ R_{r}\right\} $ are the recovery operators. The first block
in the RHS of (\ref{1}) acts on the code space $\mathcal{C}$ while the
matrices $B_{rk}$ act on $\mathcal{C}^{\perp }$ where $\mathcal{H}=$ $%
\mathcal{C}\oplus \mathcal{C}^{\perp }$. From (\ref{3}) and (\ref{1}), it
follows that DFS can be viewed as a special class of QECCs, where upon
restriction to the code space $\mathcal{C}$, all recovery operators $R_{r}$
are proportional to the inverse of the system $\mathcal{Q}$ evolution
operator, $R_{r}\propto \left( U_{\mathcal{Q}}^{\left( \text{DFS}\right)
}\right) ^{\dagger }$. Assuming that $R_{r}$ is proportional to the dagger
of $U_{\mathcal{Q}}^{\left( \text{DFS}\right) }$ , from (\ref{3}) and (\ref%
{2}) it also turns out that $A_{k}\propto U_{\mathcal{Q}}^{\left( \text{DFS}%
\right) }$ upon restriction to $\mathcal{C}$. Furthermore, from (\ref{3})
and (\ref{2}), it follows that $\gamma _{kk^{\prime }}=g_{k}^{\ast
}g_{k^{\prime }}$. However, while in the QECCs case $\gamma _{kk^{\prime }}$
is in general a full-rank matrix (non-degenerate code), in the DFSs case
this matrix has rank $1$. In conclusion, a DFS can be viewed as a special
type of QECC, namely a completely degenerate quantum error correcting code
where upon restriction to the code subspace all recovery operators are
proportional to the inverse of the system evolution operator. As a side
remark, in view of this last observation we point out that it is 
reasonable to quantify the performance of both active and passive QEC
schemes by means of the same performance measure. In what follows, we will
briefly describe the conceptual and operational meanings of the entanglement
fidelity and code entropy, respectively.

\subsection{On code performance quantifiers: entanglement fidelity and code
entropy}

\subsubsection{Entanglement Fidelity}

Entanglement fidelity is a useful measure of the efficiency of QECCs. 
It is a quantity that keeps track of how well the state and entanglement
of a subsystem of a larger system are stored, without requiring the
knowledge of the complete state or dynamics of the larger system. More
precisely, the entanglement fidelity is defined for a mixed state $\rho
=\sum_{i}p_{i}\rho _{i}=$tr$_{\mathcal{H}_{R}}\left\vert \psi \right\rangle
\left\langle \psi \right\vert $ in terms of a purification $\left\vert \psi
\right\rangle \in \mathcal{H}\otimes \mathcal{H}_{R}$ to a reference system $%
\mathcal{H}_{R}$. The purification $\left\vert \psi \right\rangle $ encodes
all of the information in $\rho $. Entanglement fidelity is a measure of how
well the channel $\Lambda $ preserves the entanglement of the state $%
\mathcal{H}$ with its reference system $\mathcal{H}_{R}$. The entanglement
fidelity is formally defined as follows \cite{schumi},%
\begin{equation}
\mathcal{F}\left( \rho \text{, }\Lambda \right) \overset{\text{def}}{=}%
\left\langle \psi |\left( \Lambda \otimes I_{\mathcal{H}_{R}}\right) \left(
\left\vert \psi \right\rangle \left\langle \psi \right\vert \right) |\psi
\right\rangle \text{,}  \label{EF}
\end{equation}%
where $\left\vert \psi \right\rangle $ is any purification of $\rho $, $I_{%
\mathcal{H}_{R}}$ is the identity map on $\mathcal{\sigma }\left( \mathcal{H}%
_{R}\right) $ and $\Lambda \otimes I_{\mathcal{H}_{R}}$ is the evolution
operator extended to the space $\mathcal{H}\otimes \mathcal{H}_{R}$, space
on which $\rho $ has been purified. If the quantum operation $\Lambda $ is
written in terms of its Kraus operator elements $\left\{ A_{k}\right\} $ as, 
$\Lambda \left( \rho \right) =\sum_{k}A_{k}\rho A_{k}^{\dagger }$, then it
can be shown that the operational expression of (\ref{EF}) becomes \cite%
{nielsen-fidel}, 
\begin{equation}
\mathcal{F}\left( \rho \text{, }\Lambda \right) =\sum_{k}\text{tr}\left(
A_{k}\rho \right) \text{tr}\left( A_{k}^{\dagger }\rho \right)
=\sum_{k}\left\vert \text{tr}\left( \rho A_{k}\right) \right\vert ^{2}\text{.%
}
\end{equation}%
This expression for the entanglement fidelity is very useful for explicit
calculations. Finally, assuming that%
\begin{equation}
\Lambda :\mathcal{\sigma }\left( \mathcal{H}\right) \ni \rho \longmapsto
\Lambda \left( \rho \right) =\sum_{k}A_{k}\rho A_{k}^{\dagger }\in \mathcal{%
\sigma }\left( \mathcal{H}\right) \text{, dim}_{%
\mathbb{C}
}\mathcal{H=}N  \label{pla1}
\end{equation}%
and choosing a purification described by a maximally entangled unit vector $%
\left\vert \psi \right\rangle \in \mathcal{H}\otimes \mathcal{H}$ for the
mixed state $\rho =\frac{1}{\text{dim}_{%
\mathbb{C}
}\mathcal{H}}I_{\mathcal{H}}$ , we obtain%
\begin{equation}
\mathcal{F}\left( \frac{1}{N}I_{\mathcal{H}}\text{, }\Lambda \right) =\frac{1%
}{N^{2}}\sum_{k}\left\vert \text{tr}A_{k}\right\vert ^{2}\text{.}
\label{nfi}
\end{equation}%
The expression\ in (\ref{nfi}) represents the entanglement fidelity when no
error correction is performed on the noisy channel $\Lambda $ in (\ref{pla1}%
).

\subsubsection{Code entropy}

A convenient quantifier of the action of the map $\Lambda $ in (\ref{channel}) 
on a initial quantum state $\rho $ is represented by the entropy
change $\mathcal{S}\left( \Lambda \left( \rho \right) \right) -\mathcal{S}%
\left( \rho \right) $. It can be shown that \cite{lind},%
\begin{equation}
0\leq \left\vert \mathcal{S}\left( \Lambda \left( \rho \right) \right) -%
\mathcal{S}\left( \sigma \right) \right\vert \leq \mathcal{S}\left( \rho
\right) \leq \mathcal{S}\left( \Lambda \left( \rho \right) \right) +\mathcal{%
S}\left( \sigma \right) \text{,}  \label{bounds}
\end{equation}%
where $\sigma =\sigma \left( \Lambda \text{, }\rho \right)$ is the
so-called Lindblad matrix, an auxiliary quantum state with matrix elements $%
\sigma _{lm}$ defined as,%
\begin{equation}
\sigma _{lm}\overset{\text{def}}{=}\text{Tr}\left( \rho A_{l}^{\dagger
}A_{m}\right) \text{,}
\end{equation}%
with $l$, $m=1$, $2$,.., $d^{2}$. The bounds provided in (\ref{bounds}) are
the so-called Lindblad bounds\emph{\ }\cite{lind}. The quantity $\mathcal{S}%
\left( \sigma \right) $ is known as the entropy exchange of the operation $%
\Lambda $. From (\ref{bounds}), it follows that if $\rho $ is pure, $%
\mathcal{S}\left( \rho \right) =0$ and therefore $\mathcal{S}\left( \sigma
\right) =\mathcal{S}\left( \Lambda \left( \rho \right) \right) $.

As pointed out earlier, a standard quantum error correction scheme for a
given quantum operation $\Lambda $ is characterized by a subspace $\mathcal{C%
}$ such that for some Hermitian and positive error correction matrix of
complex scalars $\Gamma =\left( \gamma _{lm}\right) $ the corresponding
projection operator fulfills the relation (\ref{ecc}). A quantum error
correcting code for $\Lambda $ is determined by the subspace related to $P_{%
\mathcal{C}}$. The rank of $\Gamma $ is bounded above by the dynamical Choi
matrix $\mathcal{D}_{\Lambda }$ associated with $\Lambda \left(
\rho \right) $,%
\begin{equation}
\mathcal{D}_{\Lambda }\overset{\text{def}}{=}\text{Tr}\left( A_{l}^{\dagger
}A_{m}\right) \text{.}
\end{equation}%
For orthogonal Kraus operators $\left\{ A_{l}\right\} $, the Choi matrix
equals $d_{l}\delta _{lm}$ with $0\leq d_{l}$ representing the eigenvalues
of $\mathcal{D}_{\Lambda }$. The rank of $\mathcal{D}_{\Lambda }$ equals the
minimal number of Kraus operators needed to describe $\Lambda $. Given a
code $\mathcal{C}$ for $\Lambda $ and assuming that the initial quantum
state $\rho $ belongs to the code subspace (that is, $P_{\mathcal{C}}\rho P_{%
\mathcal{C}}=\rho $), it turns out that $\sigma _{lm}=\gamma _{lm}$ \cite%
{kribs}. In other words, the error correction matrix $\Gamma $ equals the
Lindblad matrix $\sigma \left( \Lambda \text{, }\rho \right) $ provided that
the initial quantum state $\rho $ belongs to the code subspace. Motivated by
these considerations, given a quantum operation $\Lambda $ with Kraus
operators $\left\{ A_{l}\right\} $ and a code $\mathcal{C}$ with error
correction matrix $\Gamma $, Kribs and coworkers named the von Neumann
entropy $\mathcal{S}\left( \Lambda \text{, }\mathcal{C}\right) $,%
\begin{equation}
\mathcal{S}\left( \Lambda \text{, }\mathcal{C}\right) \overset{\text{def}}{=}%
\mathcal{S}\left( \Gamma \right) =-Tr\left( \Gamma \log \Gamma \right) \text{%
,}  \label{CE}
\end{equation}%
the entropy of $\mathcal{C}$ relative to $\Lambda $ \cite{kribs}. They show
that,%
\begin{equation}
0\leq \mathcal{S}\left( \Lambda \text{, }\mathcal{C}\right) \leq \log D\text{%
,}
\end{equation}%
where $\mathcal{S}\left( \Lambda \text{, }\mathcal{C}\right) =0$ iff $%
\mathcal{C}$ is a unitarily correctable code for $\Lambda $ and $\mathcal{S}%
\left( \Lambda \text{, }\mathcal{C}\right) =\log D$ iff $\mathcal{C}$ is a
non-degenerate code for $\Lambda $. From a conceptual point of view, the
entropy of a code $\mathcal{C}$ relative to a quantum noisy channel $\Lambda 
$ can be viewed as a measure quantifying the nearness of the given error
correcting code $\mathcal{C}$ to a decoherence free subspace $C_{\text{DFS}}$%
. The closer $\mathcal{C}$ is to a $C_{\text{DFS}}$, the smaller is the code
entropy $\mathcal{S}\left( \Lambda \text{, }\mathcal{C}\right) $. The
smaller is the code entropy, the smaller is the amount of effort required to
recover the quantum state corrupted by the noise by means of a suitable
error recovery operation. The simplest scenario occurs for codes with zero
entropy. Such codes are known, as we said, as unitarily correctable codes
and can be recovered with a single unitary operation. In particular
decoherence free subspaces are a special class of unitarily correctable
codes where the recovery is given by the trivial identity operation.

We recall that in standard active error correction schemes, the action of 
the recovery operation pushes all the noise into the ancillary
qubits, so that errors are eliminated when the ancilla is traced out \cite%
{laflamme-book}. That said, we emphasize that the numerical value of
the code entropy quantifies the number of ancilla qubits needed to perform a
recovery operation \cite{kribs} and the rank of $\Gamma $ gives the number
of Kraus operators necessary to describe the action of $\Lambda $ restricted
to $\mathcal{C}$ and therefore the number of Kraus operators necessary for a
recovery operation.

\section{Model I: dephasing Markovian memory channel}

In this Section, we consider a dephasing Markovian memory channel
and characterize the performance of both a repetition code and
an error avoiding code ($\mathcal{C}_{RC}$ and $\mathcal{C}_{DFS}$,
respectively) in terms of the entanglement fidelity. In particular, we
consider the concatenation of such codes ($\mathcal{C}_{DFS}\circ \mathcal{C}%
_{RC}$) and show that it is especially advantageous in the regime of partial
correlations. Finally, we characterize the effectiveness of the
codes $\mathcal{C}_{DFS}$, $\mathcal{C}_{RC}$ and $\mathcal{C}%
_{DFS}\circ \mathcal{C}_{RC}$ by means of the code entropy. We show that the
effort required for recovering such codes decreases when the error
probability decreases and the memory parameter increases.

\subsection{Entanglement fidelity-based analysis}

\emph{Repetition code for correlated phase flips}. The model considered is a 
dephasing quantum Markovian memory channel $\Lambda ^{(n)}(\rho )$.
In explicit terms, we consider $n$ qubits and Markovian correlated errors in
a dephasing quantum channel,%
\begin{equation}
\Lambda ^{(n)}(\rho )\overset{\text{def}}{=}\sum_{i_{1}\text{,..., }%
i_{n}=0}^{1}p_{i_{n}|i_{n-1}}p_{i_{n-1}|i_{n-2}}\text{...}%
p_{i_{2}|i_{1}}p_{i_{1}}\left( A_{i_{n}}\otimes \text{...}\otimes
A_{i_{1}}\right) \rho \left( A_{i_{n}}\otimes \text{...}\otimes
A_{i_{1}}\right) ^{\dagger }\text{,}  \label{n-general}
\end{equation}%
where $A_{0}\overset{\text{def}}{=}I$, $A_{1}\overset{\text{def}}{=}Z$ are
Pauli operators. Furthermore the conditional probabilities $%
p_{i_{k}|i_{k-1}}$ are given by, 
\begin{equation}
p_{i_{k}|i_{k-1}}=(1-\mu )p_{i_{k}}+\mu \delta _{i_{k}\text{, }i_{k-1}}\text{%
,}\quad p_{i_{k}=0}=1-p\text{,}\;p_{i_{k}=1}=p\text{,}  \label{aaa1}
\end{equation}%
with,%
\begin{equation}
\sum_{i_{1}\text{,..., }i_{n}=0}^{1}p_{i_{n}|i_{n-1}}p_{i_{n-1}|i_{n-2}}%
\text{...}p_{i_{2|i_{1}}}p_{i_{1}}=1\text{.}
\end{equation}%
To simplify our notation, we may choose to omit the symbol of tensor product
"$\otimes $" in the future, $A_{i_{n}}\otimes $...$\otimes A_{i_{1}}\equiv $ 
$A_{i_{n}}$...$A_{i_{1}}$. Furthermore, we may choose to omit the bar "$\mid 
$" in $p_{i_{k}|i_{j}}$ and simply write the conditional probabilities as $%
p_{i_{k}i_{j}}$. QEC is performed via the three-qubit repetition code. 
Although the error model considered is not truly quantum, we can
gain useful insights for extending error correction techniques to fully
quantum error models in the presence of partial correlations. The
performance of quantum error correcting codes is quantified by means of the
entanglement fidelity as function of the error probability $p$ and degree of
memory $\mu $. By considering the case of (\ref{n-general}) with $%
n=3$, it follows that the error superoperator $\mathcal{A}$
associated to channel is defined in terms of the following error operators,%
\begin{equation}
\mathcal{A}\longleftrightarrow \left\{ A_{0}^{\prime }\text{,.., }%
A_{7}^{\prime }\right\} \text{ with }\Lambda ^{(3)}(\rho )\overset{\text{def}%
}{=}\sum\limits_{k=0}^{7}A_{k}^{\prime }\rho A_{k}^{\prime \dagger }\text{
and, }\sum\limits_{k=0}^{7}A_{k}^{\prime \dagger }A_{k}^{\prime
}=I_{8\times 8}\text{.}  \label{nota}
\end{equation}%
In an explicit way, the error operators $\left\{ A_{0}^{\prime }\text{,.., }%
A_{7}^{\prime }\right\} $ are given by,%
\begin{eqnarray}
A_{0}^{\prime } &=&\sqrt{\tilde{p}_{0}^{\left( 3\right) }}I^{1}\otimes
I^{2}\otimes I^{3}\text{, }A_{1}^{\prime }=\sqrt{\tilde{p}_{1}^{\left(
3\right) }}Z^{1}\otimes I^{2}\otimes I^{3}\text{, }A_{2}^{\prime }=\sqrt{%
\tilde{p}_{2}^{\left( 3\right) }}I^{1}\otimes Z^{2}\otimes I^{3}\text{, } 
\notag \\
&&  \notag \\
A_{3}^{\prime } &=&\sqrt{\tilde{p}_{3}^{\left( 3\right) }}I^{1}\otimes
I^{2}\otimes Z^{3}\text{, }A_{4}^{\prime }=\sqrt{\tilde{p}_{4}^{\left(
3\right) }}Z^{1}\otimes Z^{2}\otimes I^{3}\text{, }A_{5}^{\prime }=\sqrt{%
\tilde{p}_{5}^{\left( 3\right) }}Z^{1}\otimes I^{2}\otimes Z^{3}\text{,} 
\notag \\
&&  \notag \\
\text{ }A_{6}^{\prime } &=&\sqrt{\tilde{p}_{6}^{\left( 3\right) }}%
I^{1}\otimes Z^{2}\otimes Z^{3}\text{, }A_{7}^{\prime }=\sqrt{\tilde{p}%
_{7}^{\left( 3\right) }}Z^{1}\otimes Z^{2}\otimes Z^{3}\text{,}
\end{eqnarray}%
where the coefficients $\tilde{p}_{k}^{\left( 3\right) }$ for $k=1$,.., $7$ read,
\begin{eqnarray}
\tilde{p}_{0}^{\left( 3\right) } &=&p_{00}^{2}p_{0}\text{, }\tilde{p}%
_{1}^{\left( 3\right) }=p_{00}p_{10}p_{0}\text{, }\tilde{p}_{2}^{\left(
3\right) }=p_{01}p_{10}p_{0}\text{, }\tilde{p}_{3}^{\left( 3\right)
}=p_{00}p_{01}p_{1}\text{,}  \notag \\
&&  \notag \\
\text{ }\tilde{p}_{4}^{\left( 3\right) } &=&p_{10}p_{11}p_{0}\text{, }\tilde{%
p}_{5}^{\left( 3\right) }=p_{01}p_{10}p_{1}\text{, }\tilde{p}_{6}^{\left(
3\right) }=p_{01}p_{11}p_{1}\text{, }\tilde{p}_{7}^{\left( 3\right)
}=p_{11}^{2}p_{1}\text{, }  \label{usa1}
\end{eqnarray}%
with,%
\begin{eqnarray}
p_{0} &=&\left( 1-p\right) \text{, }p_{1}=p\text{, }p_{00}=\left( \left(
1-\mu \right) \left( 1-p\right) +\mu \right) \text{, }  \notag \\
&&  \notag \\
p_{01} &=&\left( 1-\mu \right) \left( 1-p\right) \text{, }p_{10}=\left(
1-\mu \right) p\text{, }p_{11}=\left( \left( 1-\mu \right) p+\mu \right) 
\text{.}  \label{usa2}
\end{eqnarray}%
Then consider a repetition code that encodes $1$ logical qubit into 
$3$-physical qubits. The codewords are given by,%
\begin{equation}
\left\vert 0\right\rangle \rightarrow \left\vert 0_{\text{L}}\right\rangle 
\overset{\text{def}}{=}\left\vert +++\right\rangle \text{, }\left\vert
1\right\rangle \rightarrow \left\vert 1_{\text{L}}\right\rangle \overset{%
\text{def}}{=}\left\vert ---\right\rangle \text{,}  \label{placs}
\end{equation}%
where $\left\vert \pm \right\rangle \overset{\text{def}}{=}\frac{%
\left\vert 0\right\rangle \pm \left\vert 1\right\rangle }{\sqrt{2}}$. The
set of error operators satisfying the detectability condition \cite{knill02}%
, $P_{\mathcal{C}}A_{k}^{\prime }P_{\mathcal{C}}=\lambda _{A_{k}^{\prime
}}P_{\mathcal{C}}$, where $P_{\mathcal{C}}=\left\vert 0_{L}\right\rangle
\left\langle 0_{L}\right\vert +$ $\left\vert 1_{L}\right\rangle \left\langle
1_{L}\right\vert $ is the projector operator on the code subspace $\mathcal{C%
}=Span\left\{ \left\vert 0_{L}\right\rangle \text{, }\left\vert
1_{L}\right\rangle \right\} $ is given by,%
\begin{equation}
\mathcal{A}_{\text{detectable}}=\left\{ A_{0}^{\prime }\text{, }%
A_{1}^{\prime }\text{, }A_{2}^{\prime }\text{, }A_{3}^{\prime }\text{, }%
A_{4}^{\prime }\text{, }A_{5}^{\prime }\text{, }A_{6}^{\prime }\right\}
\subseteq \mathcal{A}\text{.}
\end{equation}%
The only non-detectable error is $A_{7}^{\prime }$. Furthermore, since all
the detectable errors are invertible, the set of correctable errors is such
that $\mathcal{A}_{\text{correctable}}^{\dagger }\mathcal{A}_{\text{%
correctable}}$ is detectable. It follows then that,%
\begin{equation}
\mathcal{A}_{\text{correctable}}=\left\{ A_{0}^{\prime }\text{, }%
A_{1}^{\prime }\text{, }A_{2}^{\prime }\text{, }A_{3}^{\prime }\right\}
\subseteq \mathcal{A}_{\text{detectable}}\subseteq \mathcal{A}\text{.}
\end{equation}%
The action of the correctable error operators $\mathcal{A}_{\text{correctable%
}}$ on the codewords $\left\vert 0_{L}\right\rangle $ and $\left\vert
1_{L}\right\rangle $ is given by,%
\begin{eqnarray}
\left\vert 0_{L}\right\rangle &\rightarrow &A_{0}^{\prime }\left\vert
0_{L}\right\rangle =\sqrt{\tilde{p}_{0}^{\left( 3\right) }}\left\vert
+++\right\rangle \text{, }A_{1}^{\prime }\left\vert 0_{L}\right\rangle =%
\sqrt{\tilde{p}_{1}^{\left( 3\right) }}\left\vert -++\right\rangle \text{, }%
A_{2}^{\prime }\left\vert 0_{L}\right\rangle =\sqrt{\tilde{p}_{2}^{\left(
3\right) }}\left\vert +-+\right\rangle \text{, }A_{3}^{\prime }\left\vert
0_{L}\right\rangle =\sqrt{\tilde{p}_{3}^{\left( 3\right) }}\left\vert
++-\right\rangle \text{ }  \notag \\
&&  \notag \\
\left\vert 1_{L}\right\rangle &\rightarrow &A_{0}^{\prime }\left\vert
1_{L}\right\rangle =\sqrt{\tilde{p}_{0}^{\left( 3\right) }}\left\vert
---\right\rangle \text{, }A_{1}^{\prime }\left\vert 1_{L}\right\rangle =%
\sqrt{\tilde{p}_{1}^{\left( 3\right) }}\left\vert +--\right\rangle \text{, }%
A_{2}^{\prime }\left\vert 1_{L}\right\rangle =\sqrt{\tilde{p}_{2}^{\left(
3\right) }}\left\vert -+-\right\rangle \text{, }A_{3}^{\prime }\left\vert
1_{L}\right\rangle =\sqrt{\tilde{p}_{3}^{\left( 3\right) }}\left\vert
--+\right\rangle \text{.}  \label{ea}
\end{eqnarray}%
The two \ four-dimensional orthogonal subspaces $\mathcal{V}^{0_{L}}$ and $%
\mathcal{V}^{1_{L}}$ of $\mathcal{H}_{2}^{3}$ generated by the action of $%
\mathcal{A}_{\text{correctable}}$ on $\left\vert 0_{L}\right\rangle $ and $%
\left\vert 1_{L}\right\rangle$ result,
\begin{equation}
\mathcal{V}^{0_{L}}=Span\left\{ \left\vert v_{1}^{0_{L}}\right\rangle
=\left\vert +++\right\rangle \text{,}\left\vert v_{2}^{0_{L}}\right\rangle
=\left\vert -++\right\rangle \text{, }\left\vert v_{3}^{0_{L}}\right\rangle
=\left\vert +-+\right\rangle \text{, }\left\vert v_{4}^{0_{L}}\right\rangle
=\left\vert ++-\right\rangle \text{ }\right\} \text{,}  \label{span1}
\end{equation}%
and,%
\begin{equation}
\mathcal{V}^{1_{L}}=Span\left\{ \left\vert v_{1}^{1_{L}}\right\rangle
=\left\vert ---\right\rangle \text{,}\left\vert v_{2}^{1_{L}}\right\rangle
=\left\vert +--\right\rangle \text{, }\left\vert v_{3}^{1_{L}}\right\rangle
=\left\vert -+-\right\rangle \text{, }\left\vert v_{4}^{1_{L}}\right\rangle
=\left\vert --+\right\rangle \right\} \text{,}  \label{span2}
\end{equation}%
respectively. Notice that $\mathcal{V}^{0_{L}}\oplus \mathcal{V}^{1_{L}}=%
\mathcal{H}_{2}^{3}$. The recovery superoperator $\mathcal{R}\leftrightarrow
\left\{ R_{l}\right\} $ with $l=1$,..,$4$ is defined as \cite{knill97},%
\begin{equation}
R_{l}\overset{\text{def}}{=}V_{l}\sum_{i=0}^{1}\left\vert
v_{l}^{i_{L}}\right\rangle \left\langle v_{l}^{i_{L}}\right\vert \text{,}
\label{recovery}
\end{equation}%
where the unitary operator $V_{l}$ is such that $V_{l}\left\vert
v_{l}^{i_{L}}\right\rangle =\left\vert i_{L}\right\rangle $ for $i\in
\left\{ 0\text{, }1\right\} $. Substituting (\ref{span1}) and (\ref{span2})
into (\ref{recovery}), it follows that the four recovery operators $\left\{
R_{1}\text{, }R_{2}\text{, }R_{3}\text{, }R_{4}\right\} $ are given by,%
\begin{eqnarray}
R_{1} &=&\left\vert 0_{L}\right\rangle \left\langle 0_{L}\right\vert
+\left\vert 1_{L}\right\rangle \left\langle 1_{L}\right\vert \text{, }%
R_{2}=\left\vert 0_{L}\right\rangle \left\langle -++\right\vert +\left\vert
1_{L}\right\rangle \left\langle +--\right\vert \text{,}  \notag \\
&&  \notag \\
\text{ }R_{3} &=&\left\vert 0_{L}\right\rangle \left\langle +-+\right\vert
+\left\vert 1_{L}\right\rangle \left\langle -+-\right\vert \text{, }%
R_{4}=\left\vert 0_{L}\right\rangle \left\langle ++-\right\vert +\left\vert
1_{L}\right\rangle \left\langle --+\right\vert \text{.}  \label{rec}
\end{eqnarray}%
Using simple algebra, it turns out that the $8\times 8$ matrix
representation $\left[ R_{l}\right] $ with $l=1$,..,$4$ of the recovery
operators is given by,%
\begin{equation}
\left[ R_{1}\right] =E_{11}+E_{88}\text{, }\left[ R_{2}\right] =E_{12}+E_{87}%
\text{, }\left[ R_{3}\right] =E_{13}+E_{86}\text{, }\left[ R_{4}\right]
=E_{14}+E_{85}\text{, }
\end{equation}%
where $E_{ij}$ is the $8\times 8$ matrix where the only non-vanishing
element is the one located in the $ij$-position and it equals $1$.
The action of this recovery operation $\mathcal{R}$ on the map $\Lambda
^{\left( 3\right) }\left( \rho \right) $ in (\ref{nota}) leads to,%
\begin{equation}
\Lambda _{\text{recover}}^{\left( 3\right) }\left( \rho \right) \equiv
\left( \mathcal{R\circ }\Lambda ^{(3)}\right) \left( \rho \right) \overset{%
\text{def}}{=}\sum_{k=0}^{7}\sum\limits_{l=1}^{4}\left( R_{l}A_{k}^{\prime
}\right) \rho \left( R_{l}A_{k}^{\prime }\right) ^{\dagger }\text{.}
\label{pla2}
\end{equation}%
We want to describe the action of $\mathcal{R\circ }\Lambda ^{(3)}$
restricted to the code subspace $\mathcal{C}$. Therefore, we compute the $%
2\times 2$ matrix representation $\left[ R_{l}A_{k}^{\prime }\right] _{|%
\mathcal{C}}$ of each $R_{l}A_{k}^{\prime }$ with $l=1$,.., $4$ and $k=0$%
,.., $7$ where,%
\begin{equation}
\left[ R_{l}A_{k}^{\prime }\right] _{|\mathcal{C}}\overset{\text{def}}{=}%
\left( 
\begin{array}{cc}
\left\langle 0_{L}|R_{l}A_{k}^{\prime }|0_{L}\right\rangle & \left\langle
0_{L}|R_{l}A_{k}^{\prime }|1_{L}\right\rangle \\ 
\left\langle 1_{L}|R_{l}A_{k}^{\prime }|0_{L}\right\rangle & \left\langle
1_{L}|R_{l}A_{k}^{\prime }|1_{L}\right\rangle%
\end{array}%
\right) \text{.}  \label{sopra1}
\end{equation}%
Substituting (\ref{ea}) and (\ref{rec}) into (\ref{sopra1}), it turns out
that the only matrices $\left[ R_{l}A_{k}^{\prime }\right] _{|\mathcal{C}}$
with non-vanishing trace are given by,%
\begin{eqnarray}
\left[ R_{1}A_{0}^{\prime }\right] _{|\mathcal{C}} &=&\sqrt{\tilde{p}%
_{0}^{\left( 3\right) }}\left( 
\begin{array}{cc}
1 & 0 \\ 
0 & 1%
\end{array}%
\right) \text{, }\left[ R_{2}A_{1}^{\prime }\right] _{|\mathcal{C}}=\sqrt{%
\tilde{p}_{1}^{\left( 3\right) }}\left( 
\begin{array}{cc}
1 & 0 \\ 
0 & 1%
\end{array}%
\right) \text{,}  \notag \\
&&  \notag \\
\text{ }\left[ R_{3}A_{2}^{\prime }\right] _{|\mathcal{C}} &=&\sqrt{\tilde{p}%
_{2}^{\left( 3\right) }}\left( 
\begin{array}{cc}
1 & 0 \\ 
0 & 1%
\end{array}%
\right) \text{, }\left[ R_{4}A_{3}^{\prime }\right] _{|\mathcal{C}}=\sqrt{%
\tilde{p}_{3}^{\left( 3\right) }}\left( 
\begin{array}{cc}
1 & 0 \\ 
0 & 1%
\end{array}%
\right) \text{.}
\end{eqnarray}%
Therefore, the entanglement fidelity $\mathcal{F}_{\text{phase}}^{\left(
3\right) }\left( \mu \text{, }p\right) $ defined as,%
\begin{equation}
\mathcal{F}_{\text{phase}}^{\left( 3\right) }\left( \mu \text{, }p\right) 
\overset{\text{def}}{=}\mathcal{F}^{\left( 3\right) }\left( \frac{1}{2}%
I_{2\times 2}\text{, }\mathcal{R\circ }\Lambda ^{(3)}\right) =\frac{1}{%
\left( 2\right) ^{2}}\sum_{k=0}^{7}\sum\limits_{l=1}^{4}\left\vert \text{tr}%
\left( \left[ R_{l}A_{k}^{\prime }\right] _{|\mathcal{C}}\right) \right\vert
^{2}\text{,}  \label{fidel}
\end{equation}%
results,
\begin{equation}
\mathcal{F}_{\text{phase}}^{\left( 3\right) }\left( \mu \text{, }p\right)
=\mu ^{2}\left( 2p^{3}-3p^{2}+p\right) +\mu \left( -4p^{3}+6p^{2}-2p\right)
+\left( 2p^{3}-3p^{2}+1\right) \text{.}  \label{usa3}
\end{equation}

\emph{DFS for correlated phase flips}. By considering the case of 
(\ref{n-general}) with $n=2$ qubits it follows that the
error superoperator $\mathcal{A}$ associated to channel is defined in terms
of the following error operators,%
\begin{equation}
\mathcal{A}\longleftrightarrow \left\{ A_{0}^{\prime }\text{,.., }%
A_{3}^{\prime }\right\} \text{ with }\Lambda ^{(2)}(\rho )\overset{\text{def}%
}{=}\sum\limits_{k=0}^{3}A_{k}^{\prime }\rho A_{k}^{\prime \dagger }\text{
and, }\sum\limits_{k=0}^{3}A_{k}^{\prime \dagger }A_{k}^{\prime
}=I_{4\times 4}\text{.}  \label{not}
\end{equation}%
In an explicit way, the error operators $\left\{ A_{0}^{\prime }\text{,.., }%
A_{3}^{\prime }\right\} $ are given by,%
\begin{equation}
A_{0}^{\prime }=\sqrt{\tilde{p}_{0}^{\left( 2\right) }}I^{1}\otimes I^{2}%
\text{, }A_{1}^{\prime }=\sqrt{\tilde{p}_{1}^{\left( 2\right) }}Z^{1}\otimes
I^{2}\text{, }A_{2}^{\prime }=\sqrt{\tilde{p}_{2}^{\left( 2\right) }}%
I^{1}\otimes Z^{2}\text{, }A_{3}^{\prime }=\sqrt{\tilde{p}_{3}^{\left(
2\right) }}Z^{1}\otimes Z^{2}\text{, }  \label{1a}
\end{equation}%
where the coefficients $\tilde{p}_{k}^{\left( 2\right) }$ for $k=0$,.., $3$ 
read,
\begin{equation}
\tilde{p}_{0}^{\left( 2\right) }=p_{00}p_{0}\text{, }\tilde{p}_{1}^{\left(
2\right) }=p_{10}p_{0}\text{, }\tilde{p}_{2}^{\left( 2\right) }=p_{01}p_{1}%
\text{, }\tilde{p}_{3}^{\left( 2\right) }=p_{11}p_{1}\text{.}  \label{caz}
\end{equation}%
We encode our logical qubit with a simple decoherence free subspace of two
qubits given by \cite{lidar1},%
\begin{equation}
\left\vert 0\right\rangle \longrightarrow \left\vert 0_{L}\right\rangle 
\overset{\text{def}}{=}\left\vert 01\right\rangle \text{ and, }\left\vert
1\right\rangle \longrightarrow \left\vert 1_{L}\right\rangle \overset{\text{%
def}}{=}\left\vert 10\right\rangle \text{.}  \label{pac}
\end{equation}%
The set of error operators satisfying the detectability condition \cite%
{knill02}, $P_{\mathcal{C}}A_{k}^{\prime }P_{\mathcal{C}}=\lambda
_{A_{k}^{\prime }}P_{\mathcal{C}}$, where $P_{\mathcal{C}}=\left\vert
0_{L}\right\rangle \left\langle 0_{L}\right\vert +$ $\left\vert
1_{L}\right\rangle \left\langle 1_{L}\right\vert $ is the projector operator
on the code subspace $\mathcal{C}=Span\left\{ \left\vert 0_{L}\right\rangle 
\text{, }\left\vert 1_{L}\right\rangle \right\} $ is given by,%
\begin{equation}
\mathcal{A}_{\text{detectable}}=\left\{ A_{0}^{\prime }\text{, }%
A_{3}^{\prime }\right\} \subseteq \mathcal{A}\text{.}
\end{equation}%
Furthermore, since all the detectable errors are invertible, the set of
correctable errors is such that $\mathcal{A}_{\text{correctable}}^{\dagger }%
\mathcal{A}_{\text{correctable}}$ is detectable. It follows then that,%
\begin{equation}
\mathcal{A}_{\text{correctable}}=\mathcal{A}_{\text{detectable}}\subseteq 
\mathcal{A}\text{.}
\end{equation}%
The action of the correctable error operators $\mathcal{A}_{\text{correctable%
}}$ on the codewords $\left\vert 0_{L}\right\rangle $ and $\left\vert
1_{L}\right\rangle $ is given by,%
\begin{equation}
\left\vert 0_{L}\right\rangle \rightarrow A_{0}^{\prime }\left\vert
0_{L}\right\rangle =\sqrt{\tilde{p}_{0}^{\left( 2\right) }}\left\vert
01\right\rangle \text{, }A_{3}^{\prime }\left\vert 0_{L}\right\rangle =-%
\sqrt{\tilde{p}_{3}^{\left( 2\right) }}\left\vert 01\right\rangle \text{, }%
\left\vert 1_{L}\right\rangle \rightarrow A_{0}^{\prime }\left\vert
1_{L}\right\rangle =\sqrt{\tilde{p}_{0}^{\left( 2\right) }}\left\vert
10\right\rangle \text{, }A_{3}^{\prime }\left\vert 1_{L}\right\rangle =-%
\sqrt{\tilde{p}_{3}^{\left( 2\right) }}\left\vert 10\right\rangle \text{.}
\label{ea1}
\end{equation}%
The two \ one-dimensional orthogonal subspaces $\mathcal{V}^{0_{L}}$ and $%
\mathcal{V}^{1_{L}}$ of $\mathcal{H}_{2}^{2}$ generated by the action of $%
\mathcal{A}_{\text{correctable}}$ on $\left\vert 0_{L}\right\rangle $ and $%
\left\vert 1_{L}\right\rangle $ are given by,%
\begin{equation}
\mathcal{V}^{0_{L}}=Span\left\{ \left\vert v_{1}^{0_{L}}\right\rangle
=\left\vert 01\right\rangle \right\} \text{ and, }\mathcal{V}%
^{1_{L}}=Span\left\{ \left\vert v_{1}^{1_{L}}\right\rangle =\left\vert
10\right\rangle \right\} \text{. }
\end{equation}%
Notice that $\mathcal{V}^{0_{L}}\oplus \mathcal{V}^{1_{L}}\neq \mathcal{H}%
_{2}^{2}$. This means that the trace preserving recovery superoperator $%
\mathcal{R}$ is defined in terms of one standard recovery operator $R_{1}$
and by the projector $R_{\perp }$ onto the orthogonal complement of $%
\bigoplus\limits_{i=0}^{1}\ \mathcal{V}^{i_{L}}$, i. e. the part of the
Hilbert space $\mathcal{H}_{2}^{2}$ which is not reached by acting on the
code $\mathcal{C}\ $\ with the correctable error operators. In the case
under consideration,%
\begin{equation}
R_{1}\overset{\text{def}}{=}\left\vert 01\right\rangle \left\langle
01\right\vert +\left\vert 10\right\rangle \left\langle 10\right\vert \text{, 
}R_{\perp }=\sum_{s=1}^{2}\left\vert r_{s}\right\rangle \left\langle
r_{s}\right\vert \text{,}  \label{dfsr}
\end{equation}%
where $\left\{ \left\vert r_{s}\right\rangle \right\} $ is an orthonormal
basis for $\left( \mathcal{V}^{0_{L}}\oplus \mathcal{V}^{1_{L}}\right)
^{\perp }$. A suitable basis $\mathcal{B}_{\left( \mathcal{V}^{0_{L}}\oplus 
\mathcal{V}^{1_{L}}\right) ^{\perp }}$ is given by,%
\begin{equation}
\mathcal{B}_{\left( \mathcal{V}^{0_{L}}\oplus \mathcal{V}^{1_{L}}\right)
^{\perp }}=\left\{ r_{1}=\left\vert 00\right\rangle \text{, }%
r_{2}=\left\vert 11\right\rangle \right\} \text{.}
\end{equation}%
The action of this recovery operation $\mathcal{R}$ with $%
R_{2}\equiv R_{\perp }$ on the map $\Lambda ^{\left( 2\right) }\left( \rho
\right) $ in (\ref{not}) yields,%
\begin{equation}
\Lambda _{\text{recover}}^{\left( 2\right) }\left( \rho \right) \equiv
\left( \mathcal{R\circ }\Lambda ^{(2)}\right) \left( \rho \right) \overset{%
\text{def}}{=}\sum_{k=0}^{3}\sum\limits_{l=1}^{2}\left( R_{l}A_{k}^{\prime
}\right) \rho \left( R_{l}A_{k}^{\prime }\right) ^{\dagger }\text{.}
\label{pla4}
\end{equation}%
We want to describe the action of $\mathcal{R\circ }\Lambda ^{(2)}$
restricted to the code subspace $\mathcal{C}$. Therefore, we compute the $%
2\times 2$ matrix representation $\left[ R_{l}A_{k}^{\prime }\right] _{|%
\mathcal{C}}$ of each $R_{l}A_{k}^{\prime }$ with $l=1$, $2$ and $k=0$,.., $3
$ where,%
\begin{equation}
\left[ R_{l}A_{k}^{\prime }\right] _{|\mathcal{C}}\overset{\text{def}}{=}%
\left( 
\begin{array}{cc}
\left\langle 0_{L}|R_{l}A_{k}^{\prime }|0_{L}\right\rangle  & \left\langle
0_{L}|R_{l}A_{k}^{\prime }|1_{L}\right\rangle  \\ 
\left\langle 1_{L}|R_{l}A_{k}^{\prime }|0_{L}\right\rangle  & \left\langle
1_{L}|R_{l}A_{k}^{\prime }|1_{L}\right\rangle 
\end{array}%
\right) \text{.}  \label{sup1}
\end{equation}
\begin{figure}
\centering
\includegraphics[width=0.4\textwidth]{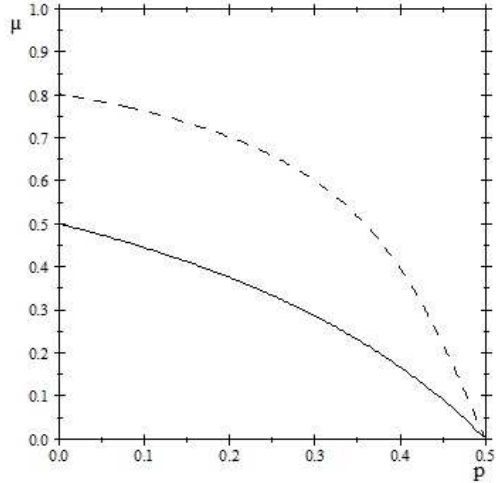}
\caption{Threshold curves for code
effectiveness: concatenated code (dashed line), DFS (solid line). The
concatenated code works in the parametric region below the dashed line. The
DFS works in the region above the solid line. The repetition code works for
any value of $\protect\mu $ when $p\leq 0.5$.
} \label{fig_1}
\end{figure}

Substituting (\ref{1a}) and (\ref%
{dfsr}) into (\ref{sup1}), it turns out that the only matrices $\left[
R_{l}A_{k}^{\prime }\right] _{|\mathcal{C}}$ with non-vanishing trace are
given by,%
\begin{equation}
\left[ R_{1}A_{0}^{\prime }\right] _{|\mathcal{C}}=\sqrt{\tilde{p}%
_{0}^{\left( 2\right) }}\left( 
\begin{array}{cc}
1 & 0 \\ 
0 & 1%
\end{array}%
\right) \text{, }\left[ R_{1}A_{3}^{\prime }\right] _{|\mathcal{C}}=-\sqrt{%
\tilde{p}_{3}^{\left( 2\right) }}\left( 
\begin{array}{cc}
1 & 0 \\ 
0 & 1%
\end{array}%
\right) 
\end{equation}%
Therefore, the entanglement fidelity $\mathcal{F}_{DFS}^{\left(
2\right) }\left( \mu \text{, }p\right) $ defined in (\ref{nfi})
results,
\begin{equation}
\mathcal{F}_{DFS}^{\left( 2\right) }\left( \mu \text{, }p\right) =\mu \left(
-2p^{2}+2p\right) +\left( 2p^{2}-2p+1\right) \text{.}  \label{dfsusa}
\end{equation}%
We point out that error correction schemes improve the transmission accuracy
only if the failure probability $\mathcal{P}\left( \mu \text{, }p\right) $
is strictly less than the error probability $p$ \cite{gaitan},%
\begin{equation}
\mathcal{P}\left( \mu \text{, }p\right) \overset{\text{def}}{=}1-\mathcal{F}%
\left( \mu \text{, }p\right) <p\text{.}  \label{curve}
\end{equation}%
The threshold curves for code effectiveness $\bar{\mu}\left(
p\right) $ are defined by the relation $1-\mathcal{F}%
\left( \bar{\mu}\left( p\right) \text{, }p\right) =0$. They allow to
select the two-dimensional parametric region where error correction schemes
are useful. The threshold curve for the DFS for the model appears in Figure \ref{fig_1}. 
The DFS works only in the parametric region above the thin solid line
in Figure \ref{fig_1}, while the repetition code works for all values of
the memory degree $\mu $ when the error probability $p$ is less than $0.5$.

\emph{Concatenated Code}. Following the line of reasoning presented in 
\cite{carloIJQI}, we consider the concatenation of the two codes just
described. In the case of (\ref{n-general}) with $n=6$ qubits and
correlated errors in a dephasing quantum channel,%
\begin{equation}
\Lambda ^{(6)}(\rho )\overset{\text{def}}{=}\sum_{i_{1}\text{,..., }%
i_{6}=0}^{1}p_{i_{6}|i_{5}}p_{i_{5}|i_{4}}p_{i_{4}|i_{3}}p_{i_{3}|i_{2}}p_{i_{2}|i_{1}}p_{i_{1}}\left( A_{i_{6}}A_{i_{5}}A_{i_{4}}A_{i_{3}}A_{i_{2}}A_{i_{1}}\right) \rho \left( A_{i_{6}}A_{i_{5}}A_{i_{4}}A_{i_{3}}A_{i_{2}}A_{i_{1}}\right) ^{\dagger }%
\text{,}  \label{rome}
\end{equation}%
The error superoperator $\mathcal{A}$ associated to channel (\ref{rome}) is
defined in terms of the following error operators,%
\begin{equation}
\mathcal{A}\longleftrightarrow \left\{ A_{0}^{\prime }\text{,.., }%
A_{63}^{\prime }\right\} \text{ with }\Lambda ^{(6)}(\rho )\overset{\text{def%
}}{=}\sum\limits_{k=0}^{2^{6}-1}A_{k}^{\prime }\rho A_{k}^{\prime \dagger }%
\text{ and, }\sum\limits_{k=0}^{2^{6}-1}A_{k}^{\prime \dagger
}A_{k}^{\prime }=I_{64\times 64}\text{.}  \label{KR}
\end{equation}%
The error operators in the Kraus decomposition (\ref{KR}) are 
$\sum_{k=0}^{6}\binom{6}{k}=2^{6}=64$, where $\binom{6}{k}$ is the
cardinality of weight-$k$ error operators.

We encode our logical qubit with a concatenated subspace obtained by
combining the decoherence free subspace in (\ref{pac}) (inner code, $%
\mathcal{C}_{DFS}=\mathcal{C}_{\text{inner}}$ ) with the repetition
code in (\ref{placs}) (outer code, $\mathcal{C}_{\text{phase}}=\mathcal{C}_{%
\text{outer}}$). We obtain that the codewords of the concatenated code $%
\mathcal{C}=\mathcal{C}_{DFS}\circ \mathcal{C}_{\text{phase}}$ are given by,%
\begin{equation}
\left\vert 0_{L}\right\rangle \overset{\text{def}}{=}\left\vert
+++---\right\rangle \text{, }\left\vert 1_{L}\right\rangle \overset{\text{def%
}}{=}\left\vert ---+++\right\rangle \text{.}  \label{CW}
\end{equation}%
Recall that the detectability condition is given by $P_{\mathcal{C}%
}A_{k}^{\prime }P_{\mathcal{C}}=\lambda _{A_{k}^{\prime }}P_{\mathcal{C}}$
where the projector operator on the code space $\mathcal{C}$ is $P_{\mathcal{%
C}}=\left\vert 0_{L}\right\rangle \left\langle 0_{L}\right\vert +\left\vert
1_{L}\right\rangle \left\langle 1_{L}\right\vert $. Observe that,%
\begin{equation}
P_{\mathcal{C}}A_{k}^{\prime }P_{\mathcal{C}}=\left\langle
0_{L}|A_{k}^{\prime }|0_{L}\right\rangle \left\vert 0_{L}\right\rangle
\left\langle 0_{L}\right\vert +\left\langle 0_{L}|A_{k}^{\prime
}|1_{L}\right\rangle \left\vert 0_{L}\right\rangle \left\langle
1_{L}\right\vert +\left\langle 1_{L}|A_{k}^{\prime }|0_{L}\right\rangle
\left\vert 1_{L}\right\rangle \left\langle 0_{L}\right\vert +\left\langle
1_{L}|A_{k}^{\prime }|1_{L}\right\rangle \left\vert 1_{L}\right\rangle
\left\langle 1_{L}\right\vert \text{.}
\end{equation}%
Therefore, it turns out that for detectable error operators we must have,%
\begin{equation}
\left\langle 0_{L}|A_{k}^{\prime }|0_{L}\right\rangle =\left\langle
1_{L}|A_{k}^{\prime }|1_{L}\right\rangle \text{ and, }\left\langle
0_{L}|A_{k}^{\prime }|1_{L}\right\rangle =\left\langle 1_{L}|A_{k}^{\prime
}|0_{L}\right\rangle =0\text{.}
\end{equation}%
In the case under consideration, it follows that the only error operator
(omitting for the sake of simplicity the proper error amplitudes) not
fulfilling the above conditions is proportional to,%
\begin{equation}
Z^{1}Z^{2}Z^{3}Z^{4}Z^{5}Z^{6}\text{.}
\end{equation}%
For such operator, we get%
\begin{equation}
\left\langle 0_{L}|Z^{1}Z^{2}Z^{3}Z^{4}Z^{5}Z^{6}|1_{L}\right\rangle =1\neq 0%
\text{ and, }\left\langle
1_{L}|Z^{1}Z^{2}Z^{3}Z^{4}Z^{5}Z^{6}|0_{L}\right\rangle =1\neq 0\text{ .}
\end{equation}%
Therefore $Z^{1}Z^{2}Z^{3}Z^{4}Z^{5}Z^{6}$ is not detectable. Thus, the
cardinality of the set of detectable errors $\mathcal{A}_{\text{detectable}}$
is $63$. Furthermore, recall that the set of correctable errors $\mathcal{A}%
_{\text{correctable}}$ is such that $\mathcal{A}_{\text{correctable}%
}^{\dagger }\mathcal{A}_{\text{correctable}}$ is detectable (in the
hypothesis of invertible error operators). Therefore, after some reasoning,
we conclude that the set of correctable errors is composed by $32$ error
operators. The correctable weight-$0$, $1$ and $2$ correctable error
operators are (omitting the proper error amplitudes),%
\begin{equation}
\left\{ I\right\} _{\text{weight-}0}\text{, }\left\{ Z^{1}\text{, }Z^{2}%
\text{, }Z^{3}\text{, }Z^{4}\text{, }Z^{5}\text{, }Z^{6}\right\} _{\text{%
weight-}1}\text{, }
\end{equation}%
and,%
\begin{equation}
\left\{ Z^{1}Z^{2}\text{, }Z^{1}Z^{3}\text{, }Z^{1}Z^{4}\text{, }Z^{1}Z^{5}%
\text{, }Z^{1}Z^{6}\text{, }Z^{2}Z^{3}\text{, }Z^{2}Z^{4}\text{, }Z^{2}Z^{5}%
\text{, }Z^{2}Z^{6}\text{, }Z^{3}Z^{4}\text{, }Z^{3}Z^{5}\text{, }Z^{3}Z^{6}%
\text{, }Z^{4}Z^{5}\text{, }Z^{4}Z^{6}\text{, }Z^{5}Z^{6}\right\} _{\text{%
weight-}2}\text{,}
\end{equation}%
respectively. The correctable weight-$3$ errors are,%
\begin{equation}
\left\{ Z^{1}Z^{3}Z^{5}\text{, }Z^{1}Z^{3}Z^{6}\text{, }Z^{1}Z^{4}Z^{5}\text{%
, }Z^{1}Z^{4}Z^{6}\text{, }Z^{1}Z^{5}Z^{6}\text{, }Z^{2}Z^{3}Z^{4}\text{, }%
Z^{2}Z^{3}Z^{5}\text{, }Z^{2}Z^{3}Z^{6}\text{, }Z^{2}Z^{4}Z^{5}\text{, }%
Z^{2}Z^{4}Z^{6}\right\} _{\text{weight-}3}\text{.}
\end{equation}%
There are no weight-$4$, $5$ and $6$ correctable error operators. The action
of the correctable errors on the codewords in (\ref{CW}) is such that the
Hilbert space $\mathcal{H}_{2}^{6}$ can be decomposed in two $32$%
-dimensional orthogonal subspaces $\mathcal{V}^{0_{L}}$ and $\mathcal{V}%
^{1_{L}}$. In other words, $\mathcal{H}_{2}^{6}=\mathcal{V}^{0_{L}}\oplus 
\mathcal{V}^{1_{L}}$ where%
\begin{equation}
\mathcal{V}^{0_{L}}=Span\left\{ \left\vert v_{k+1}^{0_{L}}\right\rangle =%
\frac{1}{\sqrt{\tilde{p}_{k}^{\left( 6\right) }}}A_{k}^{\prime }\left\vert
0_{L}\right\rangle \right\} \text{ and, }\mathcal{V}^{1_{L}}=Span\left\{
\left\vert v_{k+1}^{1_{L}}\right\rangle =\frac{1}{\sqrt{\tilde{p}%
_{k}^{\left( 6\right) }}}A_{k}^{\prime }\left\vert 1_{L}\right\rangle
\right\} \text{,}
\end{equation}%
with $A_{k}^{\prime }\in \mathcal{A}_{\text{correctable}}$ $\forall k=0$%
,..., $31$ (numbering the correctable error operators from $0$ to $31$).
Notice that $\left\langle v_{k}^{i_{L}}|v_{k^{\prime }}^{j_{L}}\right\rangle
=\delta _{kk^{\prime }}\delta _{ij}$, with $k$, $k^{\prime }\in \left\{ 0%
\text{,..., }31\right\} $ and $i$, $j\in \left\{ 0\text{, }1\right\} $ since,%
\begin{equation}
\left\langle v_{k}^{i_{L}}|v_{k^{\prime }}^{j_{L}}\right\rangle
=\left\langle i_{L}|\frac{A_{k-1}^{\prime \dagger }}{\sqrt{\tilde{p}%
_{k}^{\left( 6\right) }}}\frac{A_{k^{\prime }-1}^{\prime }}{\sqrt{\tilde{p}%
_{k^{\prime }}^{\left( 6\right) }}}|j_{L}\right\rangle =\frac{1}{\sqrt{%
\tilde{p}_{k}^{\left( 6\right) }\tilde{p}_{k^{\prime }}^{\left( 6\right) }}}%
\left\langle i_{L}|A_{k}^{\prime \dagger }A_{k^{\prime }}^{\prime
}|j_{L}\right\rangle =\frac{1}{\sqrt{\tilde{p}_{k}^{\left( 6\right) }\tilde{p%
}_{k^{\prime }}^{\left( 6\right) }}}\alpha _{kk^{\prime }}^{\prime }\delta
_{ij}=\delta _{kk^{\prime }}\delta _{ij}\text{,}
\end{equation}%
where we have used the fact that the square (Hermitian) matrix $\alpha
_{kk^{\prime }}^{\prime }$ equals $\sqrt{\tilde{p}_{k}^{\left( 6\right) }%
\tilde{p}_{k^{\prime }}^{\left( 6\right) }}$ $\delta _{kk^{\prime }}$. The
recovery superoperator $\mathcal{R}\leftrightarrow \left\{ R_{l}\right\} $
with $l=1$,.., $32$ is defined as \cite{knill97},%
\begin{equation}
R_{l}\overset{\text{def}}{=}V_{l}\sum_{i=0}^{1}\left\vert
v_{l}^{i_{L}}\right\rangle \left\langle v_{l}^{i_{L}}\right\vert \text{,}
\end{equation}%
where the unitary operator $V_{l}$ is such that $V_{l}\left\vert
v_{l}^{i_{L}}\right\rangle =\left\vert i_{L}\right\rangle $ for $i\in
\left\{ 0\text{, }1\right\} $. Notice that,%
\begin{equation}
R_{l}\overset{\text{def}}{=}V_{l}\sum_{i=0}^{1}\left\vert
v_{l}^{i_{L}}\right\rangle \left\langle v_{l}^{i_{L}}\right\vert =\left\vert
0_{L}\right\rangle \left\langle v_{l}^{0_{L}}\right\vert +\left\vert
1_{L}\right\rangle \left\langle v_{l}^{1_{L}}\right\vert \text{.}
\end{equation}%
It turns out that the $32$ recovery operators are given by,%
\begin{equation}
R_{l+1}=R_{1}\frac{A_{l}^{\prime }}{\sqrt{\tilde{p}_{l}^{\prime }}}=\left(
\left\vert 0_{L}\right\rangle \left\langle 0_{L}\right\vert +\left\vert
1_{L}\right\rangle \left\langle 1_{L}\right\vert \right) \frac{A_{l}^{\prime
}}{\sqrt{\tilde{p}_{l}^{\prime }}}\text{,}
\end{equation}%
with $l\in \left\{ 0\text{,..., }31\right\} $. Finally, the action
of this recovery operation $\mathcal{R}$ on the map $\Lambda ^{\left(
6\right) }\left( \rho \right) $ in (\ref{KR}) leads to,%
\begin{equation}
\Lambda _{\text{recover}}^{\left( 6\right) }\left( \rho \right) \equiv
\left( \mathcal{R\circ }\Lambda ^{(6)}\right) \left( \rho \right) \overset{%
\text{def}}{=}\sum_{k=0}^{2^{6}-1}\sum\limits_{l=1}^{32}\left(
R_{l}A_{k}^{\prime }\right) \rho \left( R_{l}A_{k}^{\prime }\right)
^{\dagger }\text{.}
\end{equation}%
We want to describe the action of $\mathcal{R\circ }\Lambda ^{(6)}$
restricted to the code subspace $\mathcal{C}$. Recalling that $A_{l}^{\prime
}=A_{l}^{\prime \dagger }$, it turns out that,%
\begin{equation}
\left\langle i_{L}|R_{l+1}A_{k}^{\prime }|j_{L}\right\rangle =\frac{1}{\sqrt{%
\tilde{p}_{l}^{\prime }}}\left\langle i_{L}|0_{L}\right\rangle \left\langle
0_{L}|A_{l}^{\prime \dagger }A_{k}^{\prime }|j_{L}\right\rangle +\frac{1}{%
\sqrt{\tilde{p}_{l}^{\prime }}}\left\langle i_{L}|1_{L}\right\rangle
\left\langle 1_{L}|A_{l}^{\prime \dagger }A_{k}^{\prime }|j_{L}\right\rangle 
\text{.}
\end{equation}%
We now need to compute the $2\times 2$ matrix representation $\left[
R_{l}A_{k}^{\prime }\right] _{|\mathcal{C}}$ of each $R_{l}A_{k}^{\prime }$
with $l=0$,.., $31$ and $k=0$,.., $2^{6}-1$ where,%
\begin{equation}
\left[ R_{l+1}A_{k}^{\prime }\right] _{|\mathcal{C}}\overset{\text{def}}{=}%
\left( 
\begin{array}{cc}
\left\langle 0_{L}|R_{l+1}A_{k}^{\prime }|0_{L}\right\rangle  & \left\langle
0_{L}|R_{l+1}A_{k}^{\prime }|1_{L}\right\rangle  \\ 
\left\langle 1_{L}|R_{l+1}A_{k}^{\prime }|0_{L}\right\rangle  & \left\langle
1_{L}|R_{l+1}A_{k}^{\prime }|1_{L}\right\rangle 
\end{array}%
\right) \text{.}
\end{equation}%
For $l$, $k=0$,.., $31$, we note that $\left[ R_{l+1}A_{k}^{\prime }\right]
_{|\mathcal{C}}$ becomes,%
\begin{equation}
\left[ R_{l+1}A_{k}^{\prime }\right] _{|\mathcal{C}}=\left( 
\begin{array}{cc}
\left\langle 0_{L}|A_{l}^{\prime \dagger }A_{k}^{\prime }|0_{L}\right\rangle 
& 0 \\ 
0 & \left\langle 1_{L}|A_{l}^{\prime \dagger }A_{k}^{\prime
}|1_{L}\right\rangle 
\end{array}%
\right) =\sqrt{\tilde{p}_{l}^{\prime }}\delta _{lk}\left( 
\begin{array}{cc}
1 & 0 \\ 
0 & 1%
\end{array}%
\right) \text{,}
\end{equation}%
while for any pair $\left( l\text{, }k\right) $ with $l$ $=0$,.., $31$ and $%
k>31$, it follows that,%
\begin{equation}
\left\langle 0_{L}|R_{l+1}A_{k}^{\prime }|0_{L}\right\rangle +\left\langle
1_{L}|R_{l+1}A_{k}^{\prime }|1_{L}\right\rangle =0\text{.}
\end{equation}%
We conclude that the only matrices $\left[ R_{l}A_{k}^{\prime }\right] _{|%
\mathcal{C}}$ with non-vanishing trace are given by $\left[
R_{l+1}A_{l}^{\prime }\right] _{|\mathcal{C}}$ with $l$ $=0$,.., $31$ where,%
\begin{equation}
\left[ R_{l+1}A_{l}^{\prime }\right] _{|\mathcal{C}}=\sqrt{\tilde{p}%
_{l}^{\prime }}\left( 
\begin{array}{cc}
1 & 0 \\ 
0 & 1%
\end{array}%
\right) \text{.}
\end{equation}%
Therefore, the entanglement fidelity $\mathcal{F}_{\text{conc}}^{\left(
6\right) }\left( \mu \text{, }p\right) $ defined in (\ref{nfi}) becomes,

\begin{figure}
\centering
\includegraphics[width=0.4\textwidth]{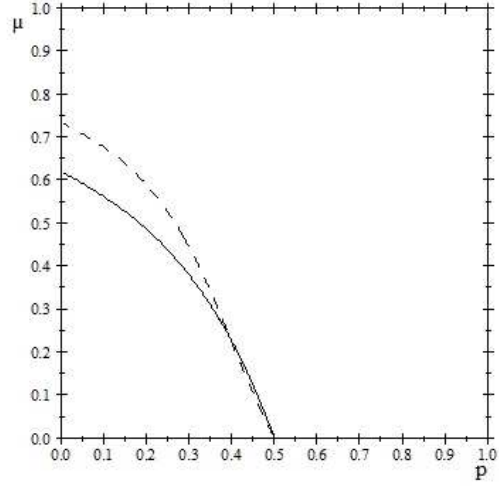}
\caption{Threshold curves for code
performance: concatenated code vs. DFS (thin solid line) and concatenated
code vs. repetition code (dashed line). The concatenated code outperforms
the DFS in the parametric region below the solid line and outperforms the
repetition code in the region below the dashed line.
} \label{fig_2}
\end{figure}

\begin{align}
\mathcal{F}_{\text{conc}}^{\left( 6\right) }\left( \mu \text{, }p\right) &
=p_{00}^{5}p_{0}+2p_{00}^{4}p_{10}p_{0}+4p_{00}^{3}p_{01}p_{10}p_{0}+p_{00}^{3}p_{10}p_{11}p_{0}+3p_{00}^{2}p_{01}p_{10}^{2}p_{0}+
\notag \\
&  \notag \\
&
+p_{00}^{3}p_{01}p_{10}p_{1}+3p_{00}^{2}p_{01}p_{10}p_{11}p_{0}+3p_{00}p_{01}^{2}p_{10}^{2}p_{0}+3p_{00}^{2}p_{01}^{2}p_{10}p_{1}+p_{00}^{3}p_{01}p_{11}p_{1}+
\notag \\
&  \notag \\
&
+p_{01}^{2}p_{10}^{3}p_{0}+2p_{00}p_{01}^{2}p_{10}^{2}p_{1}+p_{00}p_{01}p_{10}^{2}p_{11}p_{0}+p_{00}^{2}p_{01}p_{10}p_{11}p_{1}+p_{00}p_{01}p_{10}p_{11}^{2}p_{0}+
\notag \\
&  \notag \\
&
+2p_{01}^{2}p_{10}^{2}p_{11}p_{0}+p_{00}p_{01}^{2}p_{10}p_{11}p_{1}+p_{01}^{3}p_{10}^{2}p_{1}%
\text{.}
\end{align}%
Substituting (\ref{usa2}) into (\ref{EF}), we finally get%
\begin{eqnarray}
\mathcal{F}_{\text{conc}}^{\left( 6\right) }\left( \mu \text{, }p\right) 
&=&\mu ^{5}\left( -4p^{5}+\allowbreak 11p^{4}-10p^{3}+3p^{2}\right) +\mu
^{4}\left( 10p^{5}-25p^{4}+22p^{3}-8p^{2}+p\right) +  \notag \\
&&  \notag \\
&&+\mu ^{3}\left( -6p^{4}+12p^{3}-7p^{2}+p\right) +\mu ^{2}\left(
-20p^{5}+58p^{4}-60p^{3}\allowbreak +25p^{2}-3p\right) +  \notag \\
&&  \notag \\
&&+\mu \left( 20p^{5}-53p^{4}\allowbreak +46p^{3}-13p^{2}\right) +\left(
-6p^{5}+15p^{4}-10p^{3}+1\right) \text{.}  \label{CONC1}
\end{eqnarray}%
The threshold curve for code effectiveness concerning the
concatenated code defined in (\ref{CW}) for our noise model appears
in Figure \ref{fig_1}. It turns out that the concatenated code works in the
parametric region below the dashed line.

\begin{figure}
\centering
\includegraphics[width=0.4\textwidth]{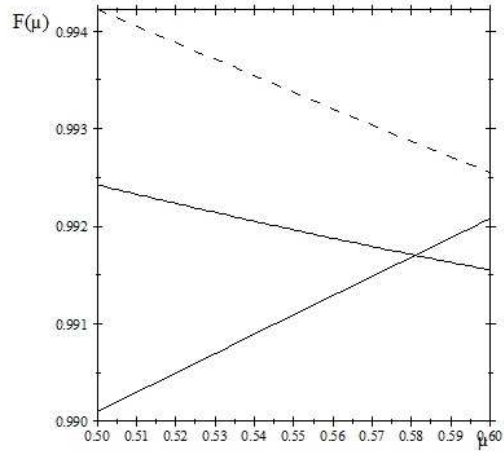}
\caption{
Entanglement fidelity vs. memory parameter $\protect\mu $ for $p=10^{-2}$:
concatenated code (dashed line), DFS (thick solid line) and repetition code
(thin solid line).} \label{fig_3}
\end{figure}

To uncover the parametric region where one code (say, code-1)
outperforms another code (say, code--2), we consider the threshold curves
for code performances $\bar{\mu}\left( p\right) $ defined by the
relation $\mathcal{F}_{\text{code-1}}\left( \bar{\mu}\left( p\right) \text{%
, }p\right) -\mathcal{F}_{\text{code-2}}\left( \bar{\mu}\left( p\right) 
\text{, }p\right) =0$. We emphasize that in view of equations (\ref{usa3}), (%
\ref{dfsusa}) and (\ref{CONC1}), it turns out that the concatenated code
outperforms the DFS in the parametric region below the solid line ($\mathcal{%
F}_{\text{conc}}^{\left( 6\right) }\left( \bar{\mu}\left( p\right) \text{, }%
p\right) -\mathcal{F}_{DFS}^{\left( 2\right) }\left( \bar{\mu}\left(
p\right) \text{, }p\right) =0$) and outperforms the three-qubit 
repetition code in the region below the dashed line ($\mathcal{F}_{\text{%
conc}}^{\left( 6\right) }\left( \bar{\mu}\left( p\right) \text{, }p\right) -%
\mathcal{F}_{\text{phase}}^{\left( 3\right) }\left( \bar{\mu}\left( p\right) 
\text{, }p\right) =0$) in Figure \ref{fig_2}. For the sake of clarity, in Figure \ref{fig_3}
we plot the entanglement fidelities (\ref{usa3}) (thin solid line), (\ref%
{CONC1}) (dashed line) and (\ref{dfsusa}) (thick solid line) for $%
p=10^{-2}$. Our analysis explicitly shows that none of the two
codes (DFS and repetition code) is effective in the extreme limit
when the other is, the repetition code still works for correlated
errors, whereas the error avoiding code does not work in the absence of
correlations. Finally, our final finding leads to conclude that
there is a parametric region characterized by intermediate values of the
memory parameter where the concatenated code in (\ref{CW}) is particularly
advantageous (see Figure \ref{fig_3}).

\subsection{Code entropy-based analysis}

\begin{figure}
\centering
\includegraphics[width=0.4\textwidth]{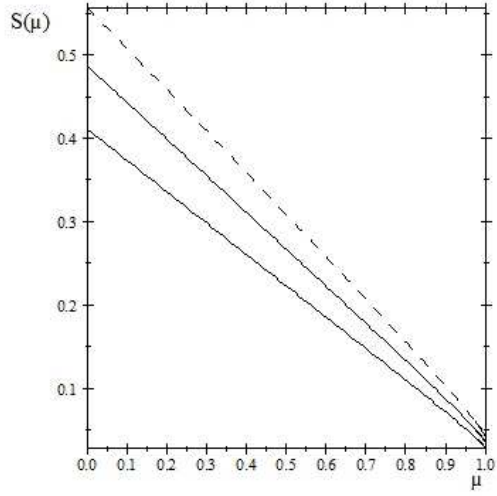}
\caption{
RC entropy vs. $\protect\mu $ for $p=2\times 10^{-2}$ (thick solid line), $p=2.5\times
10^{-2}$ (thin solid line) and $p=3\times 10^{-2}$ (dash line). The RC
entropy does not vanish for $\protect\mu =1$.
} \label{fig_4}
\end{figure}

\emph{Repetition code for correlated phase flips}. The error
correction matrix $\Gamma _{\text{phase}}$ for the three-qubit
repetition code considered is given by,%
\begin{equation}
\Gamma _{\text{phase}}=\left( 
\begin{array}{cccc}
p_{00}^{2}p_{0} & 0 & 0 & 0 \\ 
0 & p_{00}p_{10}p_{0} & 0 & 0 \\ 
0 & 0 & p_{01}p_{10}p_{0} & 0 \\ 
0 & 0 & 0 & p_{00}p_{01}p_{1}%
\end{array}%
\right) \text{.}
\end{equation}%
Therefore the repetition code entropy $\mathcal{S}_{\text{RC}%
}\left( \mu \text{, }p\right) $ results from (\ref{CE}),%
\begin{equation}
\mathcal{S}_{\text{RC}}=-p_{00}^{2}p_{0}\log _{2}\left(
p_{00}^{2}p_{0}\right) -p_{00}p_{10}p_{0}\log _{2}\left(
p_{00}p_{10}p_{0}\right) -p_{01}p_{10}p_{0}\log _{2}\left(
p_{01}p_{10}p_{0}\right) -p_{00}p_{01}p_{1}\log _{2}\left(
p_{00}p_{01}p_{1}\right) \text{,}
\end{equation}%
with,%
\begin{eqnarray}
p_{00}^{2}p_{0} &=&\mu ^{2}\left( -p^{3}+p^{2}\right) +\mu \left(
2p^{3}-4p^{2}+2p\allowbreak \right) +\left( -p^{3}+3p^{2}-3p+1\right) \text{,%
}  \notag \\
&&  \notag \\
p_{00}p_{10}p_{0} &=&p_{00}p_{01}p_{1}=\mu ^{2}\left( p^{3}-p^{2}\right)
+\mu \left( -2p^{3}+3p^{2}-p\right) +\left( p^{3}-2p^{2}+\allowbreak
p\right) \text{,}  \notag \\
&&  \notag \\
p_{01}p_{10}p_{0} &=&\mu ^{2}\left( \allowbreak p^{3}-2p^{2}+p\right) +\mu
\left( -2p^{3}+4p^{2}-\allowbreak 2p\right) +\left( p^{3}-2p^{2}+p\right) 
\text{.}
\end{eqnarray}%
In this case it turns out that there is no value of the memory parameter $%
\mu $ for which $\mathcal{C}_{\text{RC}}$ is a unitarily correctable code.
In other words, $\mathcal{S}_{\text{RC}}\left( \mu \text{, }p\right) \neq 0$
for any $0\leq \mu \leq 1$ with $0\leq p<1$. Therefore, it never occurs the
case where the effort for recovering the code is minimum. For instance, in
the extreme limit of $\mu =1$ it follows that $\mathcal{S}_{\text{RC}}\left(
\mu =1\text{, }p\right) =-\left( 1-p\right) \log _{2}\left( 1-p\right) \neq 0
$. From Figure \ref{fig_4}, we notice that in general in the limit of small error
probabilities ($p\ll 1$), the nearness of $\mathcal{C}_{RC}$ to a DFS
increases when $\mu $ increases and/or the error probability $p$ decreases.
Therefore the RC\ entropy analysis, although not particularly enlightening,
confirms that there is no pair of parametric values $\mu $ and $p$ for which 
$\mathcal{C}_{\text{RC}}$ provides a fully protected space in the presence
of correlated phase-flip errors and the effort required for recovering the
quantum state corrupted by the phase-flip noise increases when $p$
increases. 

\emph{DFS for correlated phase flips}. The error correction matrix 
$\Gamma _{\text{DFS}}$ for the DFS considered is given by,%
\begin{equation}
\Gamma _{\text{DFS}}=\left( 
\begin{array}{cc}
a & c \\ 
c & b%
\end{array}%
\right) \text{,}
\end{equation}%
where,%
\begin{equation}
a\overset{\text{def}}{=}p_{00}p_{0}\text{, }b\overset{\text{def}}{=}%
p_{11}p_{1}\text{, }c\overset{\text{def}}{=}-\sqrt{p_{00}p_{0}p_{11}p_{1}}%
\text{.}  \label{ah}
\end{equation}%
Diagonalizing $\Lambda _{\text{DFS}}$, it turns out that,%
\begin{equation}
\left[ \Lambda _{\text{DFS}}\right] _{\text{diagonal}}=\left( 
\begin{array}{cc}
\lambda _{+} & 0 \\ 
0 & \lambda _{-}%
\end{array}%
\right) \text{,}
\end{equation}%
with,%
\begin{equation}
\lambda _{\pm }=\frac{1}{2}\left[ \left( a+b\right) \pm \sqrt{\left(
a-b\right) ^{2}+4c^{2}}\right] \text{.}  \label{lam}
\end{equation}%
Substituting (\ref{ah}) into (\ref{lam}) and recalling that,%
\begin{align}
p_{0}& =\left( 1-p\right) \text{, }p_{1}=p\text{, }p_{00}=\left( \left(
1-\mu \right) \left( 1-p\right) +\mu \right) \text{, }  \notag \\
&  \notag \\
p_{01}& =\left( 1-\mu \right) \left( 1-p\right) \text{, }p_{10}=\left( 1-\mu
\right) p\text{, }p_{11}=\left( \left( 1-\mu \right) p+\mu \right) \text{,}
\end{align}%
we obtain,%
\begin{equation}
\lambda _{+}=\mu \left( -2p^{2}+2p\right) +\left( 2p^{2}-2p+1\right) \text{
and, }\lambda _{-}=0\text{.}
\end{equation}%
In conclusion the code entropy $\mathcal{S}_{\text{DFS}}\left( \mu \text{, }%
p\right) $ results from (\ref{CE}),
\begin{equation}
\mathcal{S}_{\text{DFS}}\left( \mu \text{, }p\right) =-\left[ \mu \left(
-2p^{2}+2p\right) +\left( 2p^{2}-2p+1\right) \right] \log _{2}\left[ \mu
\left( -2p^{2}+2p\right) +\left( 2p^{2}-2p+1\right) \right] \text{.}
\end{equation}

\begin{figure}
\centering
\includegraphics[width=0.4\textwidth]{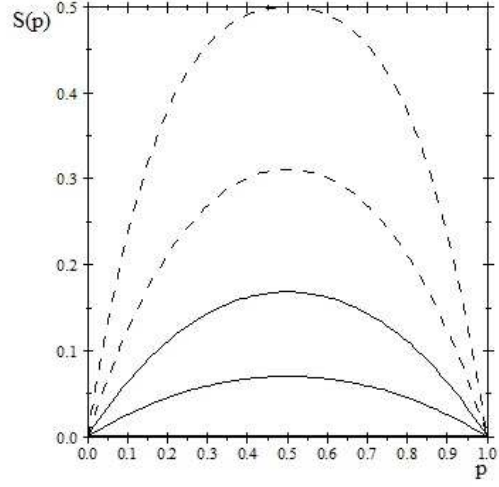}
\caption{
DFS entropy vs. $p$ for $\protect%
\mu =1$ (thick solid line), $\protect\mu =0.90$ (medium solid line), $%
\protect\mu =0.75$ (thin solid line), $\protect\mu =0.5$ (thick dash line)
and $\protect\mu =0$ (thin dash line).
} \label{fig_5}
\end{figure}

\begin{figure}
\centering
\includegraphics[width=0.4\textwidth]{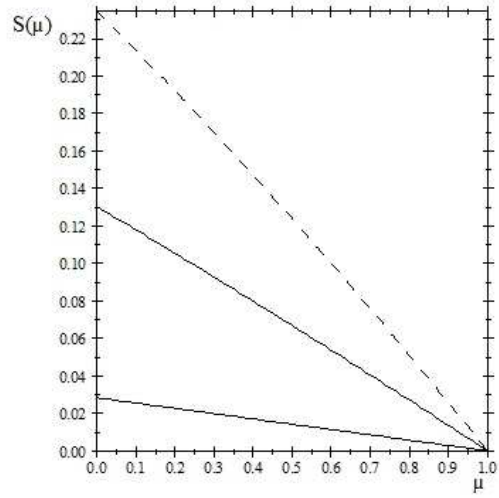}
\caption{
DFS entropy vs. memory parameter for $p=10^{-1}$ (dash solid
line), $p=5\times 10^{-2}$ (thin solid line) and $p=10^{-2}$ (medium solid
line). The DFS entropy vanishes for $\protect\mu =1$.
} \label{fig_6}
\end{figure}

From Figure \ref{fig_4}, it follows that $%
\mathcal{C}_{\text{DFS}}$ is a unitarily correctable code only in the
extreme limit of $\mu =1$. In such a case, $\mathcal{S}_{\text{DFS}}\left(
\mu =1\text{, }p\right) =0$ for any $0\leq p\leq 1$ and thus the effort for
recovering the code is minimum since a trivial identity recovery operation 
will do the job. From Figures \ref{fig_5} and \ref{fig_6}, we notice that in
general in the limit of small error probabilities ($p\ll 1$), the effort
required for recovering the quantum state corrupted by the phase-flip noise
increases when $\mu $ increases and/or $p$ increases. Thus, the DFS\ entropy
analysis leads to the conclusion that $\mathcal{C}_{\text{DFS}}$ provides an
especially useful error correction scheme for very small error
probabilities and highly correlated phase-flip errors.

\begin{figure}
\centering
\includegraphics[width=0.4\textwidth]{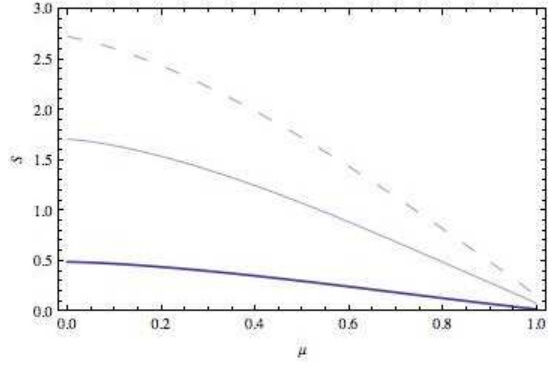}
\caption{
Concatenated code entropy vs. memory parameter $\mu$ for 
$p=10^{-1}$ (dashed solid line), $p=5\times 10^{-2}$ (thin solid line), $p=10^{-2}$ (medium solid line), 
} \label{fig_7}
\end{figure}

\emph{Concatenated Code}. After some straightforward but tedious algebra, it
turns out that the entropy of the concatenated code $\mathcal{S}_{\text{conc.%
}}\left( \mu \text{, }p\right) $ is given by,%
\begin{equation}
\mathcal{S}_{\text{conc.}}\left( \mu \text{, }p\right)
=-\sum_{j=0}^{31}f_{j}\left( \mu \text{, }p\right) \log _{2}f_{j}\left( \mu 
\text{, }p\right) \text{,}  \label{sconc}
\end{equation}%
where,%
\begin{eqnarray}
f_{0} &=&p_{00}^{5}p_{0}\text{, }f_{1}=f_{2}=p_{00}^{4}p_{10}p_{0}\text{, }%
f_{3}=f_{4}=f_{5}=f_{6}=p_{00}^{3}p_{01}p_{10}p_{0}\text{, }%
f_{7}=p_{00}^{3}p_{10}p_{11}p_{0}\text{,}  \notag \\
&&  \notag \\
f_{8} &=&f_{9}=f_{10}=p_{00}^{2}p_{01}p_{10}^{2}p_{0}\text{, }%
f_{11}=p_{00}^{3}p_{01}p_{10}p_{1}\text{, }%
f_{12}=f_{13}=f_{14}=p_{00}^{2}p_{01}p_{10}p_{11}p_{0}\text{,}  \notag \\
&&  \notag \\
f_{15} &=&f_{16}=f_{17}=p_{00}p_{01}^{2}p_{10}^{2}p_{0}\text{, }%
f_{18}=f_{19}=f_{20}=p_{00}^{2}p_{01}^{2}p_{10}p_{1}\text{, }%
f_{21}=p_{00}^{3}p_{01}p_{11}p_{1}\text{, }  \notag \\
&&  \notag \\
f_{22} &=&p_{01}^{2}p_{10}^{3}p_{0}\text{, }%
f_{23}=f_{24}=p_{00}p_{01}^{2}p_{10}^{2}p_{1}\text{, }%
f_{25}=p_{00}p_{01}p_{10}^{2}p_{11}p_{0}\text{, }%
f_{26}=p_{00}^{2}p_{01}p_{10}p_{11}p_{1}\text{,}  \notag \\
&&  \notag \\
f_{27} &=&p_{00}p_{01}p_{10}p_{11}^{2}p_{0}\text{, }%
f_{28}=f_{29}=p_{01}^{2}p_{10}^{2}p_{11}p_{0}\text{, }%
f_{30}=p_{00}p_{01}^{2}p_{10}p_{11}p_{1}\text{, }%
f_{31}=p_{01}^{3}p_{10}^{2}p_{1}\text{,}
\end{eqnarray}%
and,%
\begin{eqnarray}
p_{0} &=&\left( 1-p\right) \text{, }p_{1}=p\text{, }p_{00}=\left( \left(
1-\mu \right) \left( 1-p\right) +\mu \right) \text{, }  \notag \\
&&  \notag \\
p_{01} &=&\left( 1-\mu \right) \left( 1-p\right) \text{, }p_{10}=\left(
1-\mu \right) p\text{, }p_{11}=\left( \left( 1-\mu \right) p+\mu \right) 
\text{.}
\end{eqnarray}
It can be easily checked that in the extreme limit of $\mu =1$ it
follows that $\mathcal{S}_{\text{conc.}} \left( \mu =1\text{, }p\right)
=-\left( 1-p\right) \log _{2}\left( 1-p\right) \neq 0$. In general, it
appears that results similar to those obtained within the RC code entropy
analysis hold (see Figure \ref{fig_7}). Then, it seems that
the nearness of $\mathcal{C}_{\text{conc.}}$ to a DFS increases when $\mu $
increases and/or the error probability $p$ decreases. In particular, the
effort required for recovering the quantum state corrupted by the phase-flip
noise increases when $p$ increases.

Our analysis for Model I allows us to conclude that the entanglement
fidelity is generally a better code performance quantifier than the code
entropy since it is easier to handle computationally and, most of all, it
allows to compare performances of different error correction techniques
applied to the same error model in a smoother way. However, it appears that
the code entropy is especially informative when quantifying the performance
of DFSs in the limit of highly correlated and very low error probabilities.

\section{Model II: Asymmetric depolarizing noisy quantum memory channel}

In this Section, we consider both symmetric and asymmetric depolarizing
noisy quantum memory channels and perform quantum error correction via the
five qubit stabilizer code $\mathcal{C}_{\left[ \left[ 5,1,3\right] %
\right] }$. We characterize this code by means of the code entropy
and the entanglement fidelity as function of the error probability and the
degree of memory. In particular, we uncover that while asymmetric
depolarizing errors do not affect the entanglement fidelity of the five
qubit code, they do affect its code entropy.

\subsection{Entanglement fidelity-based analysis}

\begin{figure}
\centering
\includegraphics[width=0.4\textwidth]{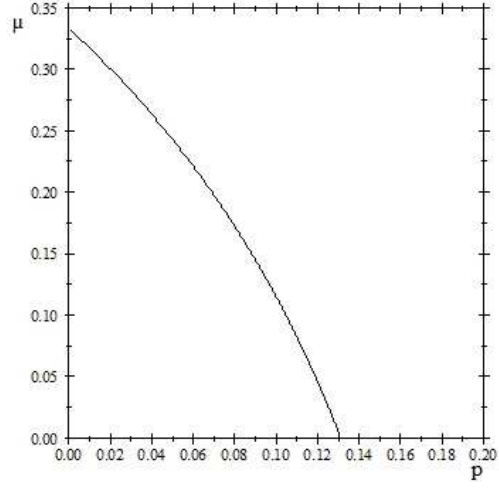}
\caption{
Threshold curve for the five-qubit stabilizer code
effectiveness. The $\mathcal{C}_{\left[ \left[ 5,1,3\right] \right] }$ code
works in the parametric region below the curve.
} \label{fig_8}
\end{figure}

\begin{figure}
\centering
\includegraphics[width=0.4\textwidth]{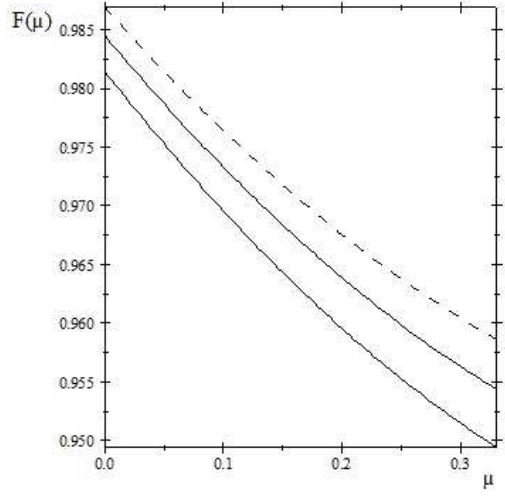}
\caption{
$\mathcal{F}^{\left[ \left[ 5,1,3\right] \right] }\left( \protect%
\mu \right) $ vs. $\protect\mu $ with $0\leq \protect\mu \leq 0.33$ ( for $%
\protect\mu >0.33$, the error correction scheme is not effective anymore)
for $p=4.50\times 10^{-2}$ (thick solid line), $p=4.10\times 10^{-2}$ (thin
solid line) and $p=3.75\times 10^{-2}$ (dashed line).
} \label{fig_9}
\end{figure}

For the symmetric case, consider five qubits and Markov correlated errors in
a depolarizing quantum channel $\Lambda ^{(5)}(\rho )$, 
\begin{equation}
\Lambda ^{(5)}(\rho )=\sum_{i_{1}\text{, }i_{2}\text{, }i_{3}\text{, }i_{4}%
\text{, }%
i_{5}=0}^{3}p_{i_{5}|i_{4}}p_{i_{4}|i_{3}}p_{i_{3}|i_{2}}p_{i_{2}|i_{1}}p_{i_{1}}%
\left[ A_{i_{5}}A_{i_{4}}A_{i_{3}}A_{i_{2}}A_{i_{1}}\rho A_{i_{1}}^{\dag
}A_{i_{2}}^{\dag }A_{i_{3}}^{\dag }A_{i_{4}}^{\dagger }A_{i_{5}}^{\dagger }%
\right] \text{,}
\end{equation}%
where $A_{0}\equiv I$, $A_{1}\equiv X$, $A_{2}\equiv Y$, $A_{3}\equiv Z$ are
the Pauli operators. The coefficients $p_{i_{l}|i_{m}}$ (conditional
probabilities) with $l$, $m$ $\in \left\{ 0\text{, }1\text{,..., }5\right\} $
satisfy the normalization condition,%
\begin{equation}
\sum_{i_{1}\text{, }i_{2}\text{, }i_{3}\text{, }i_{4}\text{, }%
i_{5}=0}^{3}p_{i_{5}|i_{4}}p_{i_{4}|i_{3}}p_{i_{3}|i_{2}}p_{i_{2}|i_{1}}p_{i_{1}}=1%
\text{.}
\end{equation}%
For the depolarizing channel $\Lambda ^{(5)}(\rho )$, coefficients $%
p_{i_{l}|i_{m}}$ are considered as, 
\begin{equation}
p_{k|k-1}\overset{\text{def}}{=}(1-\mu )p_{k}+\mu \delta _{k|k-1}\text{,}%
\quad p_{k=0}=1-p\text{,}\;p_{k=1\text{, }2\text{, }3}=p/3\text{.}
\label{conditional}
\end{equation}%
Following \cite{carloPRA}, it turns out that the entanglement fidelity $%
\mathcal{F}_{\text{symmetric}}^{\left[ \left[ 5,1,3\right] \right] }\left(
\mu \text{, }p\right) $ results from (\ref{nfi})%
\begin{equation}
\mathcal{F}_{\text{symmetric}}^{\left[ \left[ 5,1,3\right] \right] }\left(
\mu \text{, }p\right) =\sum\limits_{k=0}^{15}f_{k}\left( \mu \text{, }%
p\right) =f_{0}+\left( f_{1}+\text{...}+f_{6}\right) +\left( f_{7}+\text{...}%
+f_{15}\right) \text{,}  \label{efs}
\end{equation}%
where the functions $f_{k}\left( \mu \text{, }p\right) $ are given by,%
\begin{equation}
f_{0}=p_{00}^{4}p_{0}\text{, }f_{1}=\text{...}=f_{6}=p_{00}^{3}p_{10}p_{0}%
\text{, }f_{7}=\text{...}=f_{15}=p_{00}^{2}p_{01}p_{10}p_{0}\text{,}
\label{US1}
\end{equation}%
with,%
\begin{eqnarray}
p_{0} &=&1-p\text{, }p_{1}=\text{ }p_{2}=\text{ }p_{3}=\frac{p}{3}\text{, }%
p_{00}=\left( 1-\mu \right) \left( 1-p\right) +\mu \text{,}  \notag \\
&&  \notag \\
p_{01} &=&p_{02}=p_{03}=\left( 1-\mu \right) \left( 1-p\right) \text{, }%
p_{10}=\text{ }p_{20}=\text{ }p_{30}=\frac{p}{3}\left( 1-\mu \right) \text{.}
\end{eqnarray}%
For the asymmetric case, we assume that the error probability $p$ may be
written as,%
\begin{equation}
p=p_{X}+p_{Y}+p_{Z}\text{,}
\end{equation}%
where,%
\begin{equation}
p_{X}=\alpha _{X}p\text{, }p_{Y}=\alpha _{Y}p\text{, }p_{Z}=\alpha _{Z}p%
\text{,}
\end{equation}%
with $\alpha _{X}+\alpha _{Y}+\alpha _{Z}=1$. Notice that in the symmetric
case, we simply have $\alpha _{X}=\alpha _{Y}=\alpha _{Z}=\frac{1}{3}$. It
turns out that the $\mathcal{F}_{\text{asymmetric}}^{\left[ \left[ 5,1,3%
\right] \right] }\left( \mu \text{, }p\right) $ is given by \cite{carloPRA},%
\begin{equation}
\mathcal{F}_{\text{asymmetric}}^{\left[ \left[ 5,1,3\right] \right] }\left(
\mu \text{, }p\text{; }\alpha _{X}\text{, }\alpha _{Y}\text{, }\alpha
_{Z}\right) =\sum\limits_{k=0}^{15}f_{k}^{\prime }\left( \mu \text{, }p%
\text{; }\alpha _{X}\text{, }\alpha _{Y}\text{, }\alpha _{Z}\right)
=f_{0}^{\prime }+\left( f_{1}^{\prime }+\text{...}+f_{6}^{\prime }\right)
+\left( f_{7}^{\prime }+\text{...}+f_{15}^{\prime }\right) \text{,}
\label{d}
\end{equation}%
where the functions $f_{k}^{\prime }\left( \mu \text{, }p\text{; }\alpha _{X}%
\text{, }\alpha _{Y}\text{, }\alpha _{Z}\right) $ read,%
\begin{eqnarray}
f_{0}^{\prime } &=&p_{00}^{4}p_{0}\text{, }f_{1}^{\prime
}=p_{00}^{3}p_{0}p_{10}\text{, }f_{2}^{\prime }=p_{00}^{3}p_{0}p_{20}\text{, 
}f_{3}^{\prime }=p_{00}^{3}p_{0}p_{30}\text{, }f_{4}^{\prime
}=p_{00}^{3}p_{01}p_{1}\text{, }f_{5}^{\prime }=p_{00}^{3}p_{01}p_{2}\text{, 
}f_{6}^{\prime }=p_{00}^{3}p_{01}p_{3}\text{,}  \notag \\
&&  \notag \\
f_{7}^{\prime } &=&f_{8}^{\prime }=f_{9}^{\prime
}=p_{00}^{2}p_{01}p_{0}p_{10}\text{, }f_{10}^{\prime }=f_{11}^{\prime
}=f_{12}^{\prime }=p_{00}^{2}p_{01}p_{0}p_{20}\text{, }f_{13}^{\prime
}=f_{14}^{\prime }=f_{15}^{\prime }=p_{00}^{2}p_{01}p_{0}p_{30}\text{,}
\label{US2}
\end{eqnarray}%
with,%
\begin{eqnarray}
p_{0} &=&1-p\text{, }p_{1}=\alpha _{X}p\text{, }p_{2}=\alpha _{Y}p\text{, }%
p_{3}=\alpha _{Z}p\text{, }p_{00}=\left( 1-\mu \right) \left( 1-p\right)
+\mu \text{,}  \notag \\
&&  \notag \\
p_{01} &=&p_{02}=p_{03}=\left( 1-\mu \right) \left( 1-p\right) \text{, }%
p_{10}=\alpha _{X}p\left( 1-\mu \right) \text{, }p_{20}=\alpha _{Y}p\left(
1-\mu \right) \text{, }p_{30}=\alpha _{Z}p\left( 1-\mu \right) \text{.}
\label{SUP}
\end{eqnarray}%
Recalling that in the symmetric case $p_{1}=p_{2}=p_{3}=\frac{p}{3}$ and $%
p_{10}=p_{20}=p_{30}=\frac{p}{3}\left( 1-\mu \right) $ and substituting (\ref%
{SUP}) in (\ref{d}), it follows that%
\begin{equation}
\mathcal{F}_{\text{asymmetric}}^{\left[ \left[ 5,1,3\right] \right] }\left(
\mu \text{, }p\text{; }\alpha _{X}\text{, }\alpha _{Y}\text{, }\alpha
_{Z}\right) =\mathcal{F}_{\text{symmetric}}^{\left[ \left[ 5,1,3\right] %
\right] }\left( \mu \text{, }p\right) \text{.}  \label{equality}
\end{equation}%
Therefore, we conclude that the performance of the five-qubit stabilizer 
code, when quantified by the entanglement fidelity, remains
unaffected by the asymmetry of the depolarizing error probabilities. From (%
\ref{d}) it can be shown that such error correction scheme only works for
low values of $\mu $ (see Figure \ref{fig_8}). Furthermore, it also turns out that
the performance of the five qubit quantum stabilizer code is lowered by
increasing values of the degree of memory $\mu $ and increasing values of
the error probability values $p$ (see Figure \ref{fig_9}). For a more detailed
overview of such findings, we refer to \cite{carloPRA}.

\subsection{Code entropy-based analysis}

\begin{figure}
\centering
\includegraphics[width=0.4\textwidth]{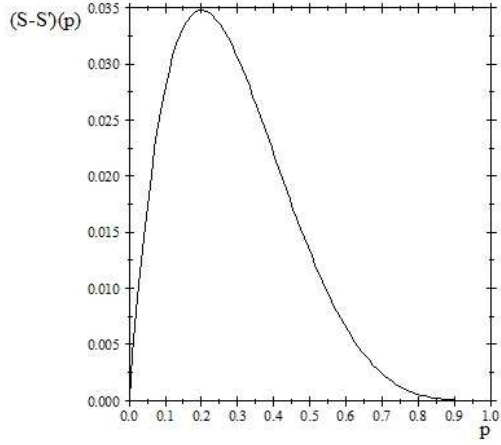}
\caption{
Difference between the five qubit
stabilizer code entropies $\mathcal{S}_{\text{symmetric}}^{\left[ \left[
5,1,3\right] \right] }\left( p\right) $ and $\mathcal{S}_{\text{asymmetric}%
}^{\left[ \left[ 5,1,3\right] \right] }\left( p\right) $ in the absence of
correlations ($\protect\mu =0$) with $\protect\alpha _{x}=\protect\alpha %
_{y}=\frac{1}{4}$ and $\protect\alpha _{z}=\frac{1}{2}$.
} \label{fig_10}
\end{figure}

Omitting technical details and following the work presented in the previous
Section, it turns out that the five qubit code entropies for the symmetric
and asymmetric depolarizing quantum memory channels are given by,%
\begin{equation}
\mathcal{S}_{\text{symmetric}}^{\left[ \left[ 5,1,3\right] \right] }\left(
\mu \text{, }p\right) =-\sum_{j=0}^{15}f_{j}\left( \mu \text{, }p\right)
\log _{2}f_{j}\left( \mu \text{, }p\right) \text{,}  \label{sym5}
\end{equation}%
and,%
\begin{equation}
\mathcal{S}_{\text{asymmetric}}^{\left[ \left[ 5,1,3\right] \right] }\left(
\mu \text{, }p\text{; }\alpha _{X}\text{, }\alpha _{Y}\text{, }\alpha
_{Z}\right) =-\sum_{j=0}^{15}f_{j}^{\prime }\left( \mu \text{, }p\text{; }%
\alpha _{X}\text{, }\alpha _{Y}\text{, }\alpha _{Z}\right) \log
_{2}f_{j}^{\prime }\left( \mu \text{, }p\text{; }\alpha _{X}\text{, }\alpha
_{Y}\text{, }\alpha _{Z}\right) \text{,}  \label{asym5}
\end{equation}%
respectively. The explicit expressions for the functions $f_{j}\left( \mu 
\text{, }p\right) $ and $f_{j}^{\prime }\left( \mu \text{, }p\text{; }\alpha
_{X}\text{, }\alpha _{Y}\text{, }\alpha _{Z}\right) $ are given in (\ref{US1}%
) and (\ref{US2}), respectively. From (\ref{sym5})\ and (\ref{asym5}), it
can be shown that 
\begin{equation}
\mathcal{S}_{\text{symmetric}}^{\left[ \left[ 5,1,3\right] \right] }\left(
\mu \text{, }p\right) \neq \mathcal{S}_{\text{asymmetric}}^{\left[ \left[
5,1,3\right] \right] }\left( \mu \text{, }p\text{; }\alpha _{X}\text{, }%
\alpha _{Y}\text{, }\alpha _{Z}\right) \text{.}  \label{dise}
\end{equation}%
Numerical evidence of the disequality in (\ref{dise}) can be easily verified
at least for a suitable choice of model parameters $\mu $, $\alpha _{X}$, $%
\alpha _{Y}$ and $\alpha _{Z}$. In Figure \ref{fig_10}, we plot the change
of the five qubit code entropy $\Delta _{\mathcal{S}}\left( p\right) $ as
function of the error probability $p$. Specifically, $\Delta _{%
\mathcal{S}}\left( p\right) $ is the difference between the five qubit code
entropies $\mathcal{S}_{\text{symmetric}}^{\left[ \left[ 5,1,3\right] \right]
}\left( p\right) $ and $\mathcal{S}_{\text{asymmetric}}^{\left[ \left[ 5,1,3%
\right] \right] }\left( p\right) $ in the absence of correlations ($\mu =0$)
with $\alpha _{x}=\alpha _{y}=\frac{1}{4}$ and $\alpha _{z}=\frac{1}{2}$.
The positivity of $\Delta _{\mathcal{S}}\left( p\right) $ leads to conclude
that when the error model considered is defined by a suitable choice of
numerical values of the model parameters, it can happen that the effort
required for recovering the five qubit code in the asymmetric case is less
than in the symmetric case. This scenario never occurs when quantifying the
performance of the five qubit code applied to asymmetric and correlated
depolarizing errors by means of the entanglement fidelity.

From an analytical point
of view, (\ref{dise}) can be readily understood by noticing that $\mathcal{S}%
_{\text{symmetric}}^{\left[ \left[ 5,1,3\right] \right] }\left( \mu \text{, }%
p\right) $ in (\ref{sym5}) and $\mathcal{S}_{\text{asymmetric}}^{\left[ %
\left[ 5,1,3\right] \right] }\left( \mu \text{, }p\text{; }\alpha _{X}\text{%
, }\alpha _{Y}\text{, }\alpha _{Z}\right) $ in (\ref{asym5}) are sums of
terms that are \emph{nonlinear} in the functions (amplitude square
probabilities of the correctable error operators) $f_{j}\left( \mu \text{, }%
p\right) $ and $f_{j}^{\prime }\left( \mu \text{, }p\text{; }\alpha _{X}%
\text{, }\alpha _{Y}\text{, }\alpha _{Z}\right) $, respectively. As
a matter of fact it turns out that,
\begin{equation}
\sum_{k=0}^{15}f_{k}=\sum_{k=0}^{15}f_{k}^{\prime }\text{,}  \label{R1}
\end{equation}%
with $f_{0}^{\prime }=f_{0}=p_{00}^{4}p_{0}$, $f_{1}^{\prime
}+...+f_{6}^{\prime }=f_{1}+...+f_{6}$, $f_{7}^{\prime }+...+f_{15}^{\prime
}=f_{7}+...+f_{15}$ and, $f_{i}^{\prime }\neq f_{j}^{\prime }$ for $i$, $j=1$%
,..., $6$; $f_{7}^{\prime }=f_{8}^{\prime }=f_{9}^{\prime
}=p_{00}^{2}p_{01}p_{0}p_{10}$; $f_{10}^{\prime }=f_{11}^{\prime
}=f_{12}^{\prime }=p_{00}^{2}p_{01}p_{0}p_{20}$ and $f_{13}^{\prime
}=f_{14}^{\prime }=f_{15}^{\prime }=p_{00}^{2}p_{01}p_{0}p_{30}$.
However, the constraint in (\ref{R1}) together with the nonlinearity in $f$ 
and $f^{\prime }$ characterizing the expressions of the
code entropies in (\ref{sym5}) and (\ref{asym5}) imply that,
\begin{equation}
\sum_{k=0}^{15}f_{k}\log _{2}f_{k}\neq \sum_{k=0}^{15}f_{k}^{\prime }\log
_{2}f_{k}^{\prime }\text{.}  \label{R2}
\end{equation}%
In conclusion, since the entanglement fidelity is a linear combination 
of the amplitude square probabilities of the correctable error 
operators, Eq.\ (\ref{R1}) holds true, and the equality in (\ref{equality}) is proven. 
On the contrary, the code entropy is a 
\emph{nonlinear} combination of the amplitude square probabilities
and the use of (\ref{R2}) leads naturally to the inequality in (\ref{dise}).

Therefore, we conclude that the performance of the five-qubit 
stabilizer code, when quantified by the code entropy and not by the
entanglement fidelity, does remain affected by the asymmetry of the
depolarizing error probabilities. Our finding provides a neat quantitative
manifestation of the conceptual fact \cite{kribs} that no single code
performance quantifier holds all the information on a code.

\section{Final Remarks}

In this article, we studied the properties of error correcting codes
for noise models in the presence of asymmetries and/or correlations by means
of the entanglement fidelity and the code entropy. We considered a
dephasing Markovian memory channel (Model I) and both symmetric and
asymmetric depolarizing quantum memory channels (Model II). For each model,
we presented both an entanglement fidelity-based and a code entropy-based
analyses. For Model I, we used three codes: the repetition code 
$\mathcal{C}_{RC}$, the DFS $\mathcal{C}_{DFS}$ and the concatenated code 
$\mathcal{C}_{DFS}\circ \mathcal{C}_{RC}$. For Model II, we employed the five
qubit stabilizer code $\mathcal{C}_{\left[ \left[ 5,1,3\right] \right] }$.

For Model I, the entanglement fidelity-based analysis allows to find out the
parametric regions where the chosen error correction schemes are effective
(see Figure \ref{fig_1}). This analysis is also suitable to determine where, within
such parametric regions, one code outperforms the other (see Figure \ref{fig_2}). In
particular, we showed that the concatenated code quantified by the
entanglement fidelity is useful to combat partially correlated
phase-flip errors (see Figure \ref{fig_3}). The code entropy-based analysis for Model
I leads to the conclusion that the effort required for recovering the
quantum state corrupted by correlated phase-flip errors increases when the
memory parameter $\mu $ decreases and the error probability $p$ increases
(see Figures \ref{fig_4} and \ref{fig_6}). 
Furthermore, it turns out that only $\mathcal{C}_{DFS}$ is a unitarily correctable 
code for the noise model considered in the limiting case of $\mu =1$ (see Figure \ref{fig_5}). 
When applied to the
concatenated code, the code entropy-based analysis is not as enlightening as
the entanglement-fidelity based analysis. It is not particularly useful for drawing
performance comparisons with the repetition code (see Figure \ref{fig_7}).

For Model II, the entanglement fidelity-based analysis allows us to
find out that the five qubit stabilizer code only works
for small values of $\mu $ (see Figure \ref{fig_8}). Furthermore, the performance of 
this code applied to symmetric and asymmetric correlated
depolarizing errors is lowered by increasing values of the memory parameter $%
\mu $ and increasing values of the error probability $p$ (see Figure \ref{fig_9}).
Finally, the code entropy-based analysis applied to Model II leads to an
interesting result. While asymmetry in the depolarizing errors does not
affect in any case the performance of the five qubit stabilizer code
quantified by means of the entanglement fidelity, it may affect positively
the performance of $\mathcal{C}_{\left[ \left[ 5,1,3\right] \right] }$ by
lowering the effort required for recovering the code subjected to asymmetric
depolarizing errors (see Figure \ref{fig_10}).

Although our work is limited to only few error models in the presence of
correlations and asymmetry and we only perform error correction by means of
few quantum codes, we feel we have gathered enough evidence to draw the
following conclusions on the code performance quantifiers employed:\ a)
while the code entropy may capture new undetected features of a code with
respect to those encoded into the entanglement fidelity and deserves further 
investigations, it is certainly harder to compute for realistic
physical models and non-trivial quantum codes; b) the code entropy seems to
lack the reliability and practicality characterizing the entanglement
fidelity.

In summary, the entropy of a code appears to be an interesting
auxiliary tool for quantifying the code performances. In
particular, it is our intention to explore in future investigations the
possibility of devising a hybrid code performance quantifier whose nonlinear
structure in the amplitude square probabilities of the correctable error
operators may not be necessarily characterized by the logarithmic behavior
which identifies the code entropy. For the time being, motivated by our
analysis and in agreement with \cite{nielsen-fidel, cory, knill-exp}, we
believe that the entanglement fidelity remains the most relevant tool to
maximize in schemes for quantum error correction.

\begin{acknowledgments}
The research leading to these results has received funding from the
European Commission's Seventh Framework Programme (FP7/2007--2013) under
grant agreements no. 213681.
\end{acknowledgments}

\end{document}